\documentclass[prd,tightenlines,nofootinbib,superscriptaddress]{revtex4}

\usepackage{amsfonts,amssymb,amsthm,bbm,amsmath}

\usepackage{color,psfrag}
\usepackage[dvips]{graphicx}

\newcommand{\C}{{\mathbb C}}
\newcommand{\N}{{\mathbb N}}
\newcommand{\R}{{\mathbb R}}

\newcommand{\cA}{{\mathcal A}}

\newcommand{\cH}{{\mathcal H}}
\newcommand{\cM}{{\mathcal M}}
\newcommand{\cN}{{\mathcal N}}
\newcommand{\cR}{{\mathcal R}}
\newcommand{\cO}{{\mathcal O}}

\newcommand{\cV}{{\mathcal V}}

\newcommand{\cC}{{\mathcal C}}
\newcommand{\cS}{{\mathcal S}}
\newcommand{\SU}{\mathrm{SU}}
\newcommand{\Spin}{\mathrm{Spin}}
\newcommand{\SL}{\mathrm{SL}}
\newcommand{\SO}{\mathrm{SO}}
\newcommand{\U}{\mathrm{U}}

\newcommand{\vJ}{\vec{J}}
\newcommand{\vK}{\vec{K}}

\newcommand{\be}{\begin{equation}}
\newcommand{\ee}{\end{equation}}
\newcommand{\beq}{\begin{eqnarray}}
\newcommand{\eeq}{\end{eqnarray}}
\newcommand{\bes}{\begin{eqnarray}}
\newcommand{\ees}{\end{eqnarray}}

\newcommand{\mat} [2] {\left ( \begin{array}{#1}#2\end{array} \right ) }

\newcommand{\su}{{\mathfrak su}}
\newcommand{\spin}{{\mathfrak spin}}
\renewcommand{\u}{{\mathfrak u}}

\newcommand{\la}{\langle}
\newcommand{\ra}{\rangle}

\newcommand{\tr}{{\mathrm{Tr}}}
\newcommand{\f}{\frac}

\newcommand{\vphi}{\varphi}

\def\bF{{\bar{F}}}

\newcommand{\bz}{\overline{z}}

\newcommand{\Ref}[1]{(\ref{#1})}

\def\nn{\nonumber}
\def\pp{\partial}
\def\te{{\widetilde{e}}}
\def\tf{{\widetilde{f}}}
\def\arr{\rightarrow}

\def\vphi{\varphi}

\def\hE{\hat{E}}
\def\hF{\hat{F}}
\def\hFd{{\hat{F}^\dag}}

\def\cHs{{\cal H}^{simple}}
\def\rpsi{{}_\rho\psi}
\def\tw{\widetilde{w}}
\def\tz{\widetilde{z}}
\def\eps{\epsilon}
\def\tdelta{\widetilde{\delta}}

\newcommand{\id}{\mathbb{I}}
\def\vV{\vec{V}}
\def\vsigma{\vec{\sigma}}

\newtheorem{theorem}{Theorem}[section]

\newtheorem{prop}[theorem]{Proposition}

\begin{document}

\title{Holomorphic Simplicity Constraints for 4d Spinfoam Models}

\author{{\bf Ma\"it\'e Dupuis}}\email{maite.dupuis@ens-lyon.fr}
\affiliation{Laboratoire de Physique, ENS Lyon, CNRS-UMR 5672, 46 All\'ee d'Italie, Lyon 69007, France}
\affiliation{School of Physics, The University of Sydney, Sydney, New South Wales 2006, Australia}
\author{{\bf Etera R. Livine}}\email{etera.livine@ens-lyon.fr}
\affiliation{Laboratoire de Physique, ENS Lyon, CNRS-UMR 5672, 46 All\'ee d'Italie, Lyon 69007, France}

\date{\today}

\begin{abstract}


Within the framework of spinfoam models, we revisit the simplicity constraints reducing topological BF theory to 4d Riemannian gravity.
We use the reformulation of $\SU(2)$ intertwiners and spin networks in terms of spinors, which has come out from both the recently developed $\U(N)$ framework for $\SU(2)$ intertwiners and the twisted geometry approach to spin networks and spinfoam boundary states.
Using these tools, we are able to perform a holomorphic/anti-holomorphic splitting of the simplicity constraints and define a new set of holomorphic simplicity constraints, which are equivalent to the standard ones at the classical level and which can be imposed strongly on intertwiners at the quantum level. We then show how to solve these new holomorphic simplicity constraints using coherent intertwiner states. We further define the corresponding coherent spin network functionals and introduce a new spinfoam model for 4d Riemannian gravity based on these holomorphic simplicity constraints and whose amplitudes are defined from the evaluation of the new coherent spin networks.

\end{abstract}

\maketitle


\section*{Introduction}

The Spinfoam framework for quantum gravity is a formalism for a regularized path integral for general relativity. It is based on a reformulation of gravity as a quasi-topological field theory, or more precisely as a topological BF theory with extra constraints which break the topological invariance and (re-)introduce local degrees of freedom in the field theory. These are called the {\it simplicity constraints} and are at the heart of the spinfoam quantization program. They correspond to the second class constraints appearing in the canonical treatment of the first order (Holst-)Palatini action for general relativity as a gauge field theory\footnotemark.
\footnotetext{
They also appear as the reality conditions in the self-dual Ashtekar formulation of general relativity as an $\SU(2)$ gauge theory.}

The standard procedure to construct a spinfoam path integral for quantum gravity is to start from the discretized path integral for the topological BF theory, either under its state-sum formulation or derived from a discrete BF action. Then one discretizes the simplicity constraints, investigates their geometrical and physical meaning at the spinfoam level and imposes them. Finally, the goal is to check whether this leads to the correct degrees of freedom both at the fundamental discrete level and in the semi-classical continuum level at large scales.

The first explicit spinfoam model for 4d quantum gravity is the Barrett-Crane model, both in its Riemannian version \cite{bc1} and Lorentzian version \cite{bc2}. It relies on a strong imposition of the discrete simplicity constraints. Since then, it has been argued that this is a too strong requirement. Indeed the strong imposition seems to kill too many degrees of freedom (e.g. \cite{graviton}). Moreover it turns out that the Hilbert space of boundary states of the Barrett-Crane model does not seem to fit with the space of canonical states of Loop Quantum Gravity, which is another inconvenience (see e.g. \cite{proj,clqg}). It was thus later argued that a weaker imposition of the (discrete) simplicity constraints would improve the spinfoam procedure and the semi-classical behavior of the model \cite{LS,EPR}. This is related to the fact that the simplicity constraints correspond to second class constraints in the canonical analysis of the classical theory. It was then proposed to solve them only weakly in the spinfoam path integral through coherent state techniques \cite{LS,FK,LS2} or through a Gupta-Bleuer-like procedure \cite{EPR,EPRL}. This lead to the definition of the EPRL-FK spinfoam models \cite{FK,EPRL} which exist for both Riemannian and Lorentzian space-time signature (also see \cite{lorentzian} for a more thorough definition of the Lorentzian spinfoam model). This is the current state-of-the-art spinfoam proposal for 4d quantum gravity.

Nevertheless, the construction of the EPRL-FK models relies on imposing the simplicity constraints on the expectation values with small uncertainty\footnotemark.
\footnotetext{
Actually, the requirement is slightly stronger and we require the vanishing of the matrix elements of the constraints between solution states. Calling $\cH_s$ the Hilbert space of solution states and $\cC$ the simplicity constraints, we ask \cite{EPR,LS2}:
$$
\forall \psi,\phi\in\cH_s,\quad\la\psi|\cC|\phi\ra=0.
$$
This is stronger than simply requiring the vanishing of the expectation values $\la\psi|\cC|\psi\ra=0$, but in practice it amounts simply to vanishing of the expectation values with small (almost-minimal) uncertainty.
}
This ensures a nice behavior of the states in the semi-classical regime. But a drawback with this construction is that the states are not properly defined as actual (strong) solutions to a set of constraints. In particular, they do not come from an actual Gupta-Bleuer procedure with an holomorphic/anti-holomorphic factorization of the constraints in terms of creation and annihilation operators. This means that we can not define the EPRL-FK states through a simple algebraic equation.

Recently in \cite{un3}, we have proposed a detailed analysis of the algebraic properties of the discrete simplicity constraints for 4d Riemannian gravity using the $\U(N)$ framework for $\SU(2)$ intertwiners \cite{un0,un1,un2}. This lead to identification of an holomorphic/anti-holomorphic factorization of the simplicity constraints, which can now be used to perform a true Gupta-Bleuer procedure in order to take into account the simplicity constraints. We can indeed use these new $F$-simplicity constraints and impose them strongly. In \cite{un3}, it was shown that this solves the simplicity constraints weakly and that we can construct solutions as coherent states using tools from the $\U(N)$ framework. In the present paper, we propose to investigate further the definition and properties of these $F$-simplicity constraints, both at the classical and quantum levels. Furthermore, we build new spinfoam amplitudes based on imposing these $F$-simplicity constraints and we compare them to the standard EPRL-FK models.


In the next section, we start by reviewing the classical phase space formulation for $\SU(2)$ intertwiners in terms of spinors. This provides a clear geometrical framework to translate the simplicity constraints. We introduce our new $F$-simplicity constraints, or  holomorphic simplicity constraints, and show their equivalence with the standard ways to formulate the simplicity constraints. We discuss the geometrical meaning of the various simplicity constraints.
In section II, we move to the quantum level and review the $\U(N)$ framework for $\SU(2)$ intertwiners, which is obtained as a quantization of the classical spinor phase space. We review how to define coherent intertwiner states in this framework. Then we show how to solve the holomorphic simplicity constraints at the quantum level using these tools.
In section III, we propose a classical action principle to define the classical dynamics of discrete geometry states solving the holomorphic simplicity constraints. In the final section, we define a new spinfoam model solving strongly the holomorphic simplicity constraints and weakly the standard simplicity constraints, and we show its link to the EPRL-FK spinfoam amplitudes.


We work in this whole paper with 4d Riemannian quantum gravity. The Lorentzian case will be investigated elsewhere \cite{Linprep}.

\section{Classical geometry of the Simplicity Constraints}

$\SU(2)$ intertwiners are the basic building blocks of the spin network states for quantum geometry in Loop Quantum Gravity. Recently, the $\U(N)$ framework for the Hilbert space of intertwiners was developed \cite{un0,un1,un2,un3,un4} and it was realized that it leads to a re-formulation of intertwiner states as the quantization of a classical phase space parameterized by spinors \cite{un2,un4}. When gluing the intertwiners back together along graphs to form spin network states, this actually led back to the twisted geometries, introduced independently to describe the discrete phase space of loop quantum gravity on a fixed graph \cite{twisted,twistor}.

In this section, we will review this classical spinor phase space formalism for $\SU(2)$-invariant states and generalize it to $\Spin(4)\sim \SU_L(2)\times\SU_R(2)$. We will then discuss the simplicity constraints in this framework: we will define our new holomorphic simplicity constraints and show their equivalence with the standard expressions of the simplicity constraints at the classical level.

\subsection{The Classical Spinor Framework for $\SU(2)$}

Following the previous ideas on the $\U(N)$ formalism for intertwiners \cite{un1,un2,un3} and on twisted geometries for loop gravity and spin foams \cite{twisted,twistor}, it was realized that loop quantum gravity's spin network states are the quantization of some classical spinor networks \cite{un4}. We review this formalism below.

Spin networks, and thus spinor networks, are constructed on a given graph. Let us thus choose a closed oriented graph $\Gamma$ with $E$ edges and $V$ vertices. We will label its vertices as $v$ and its edges as $e$, calling $s(e)$ and $t(e)$ respectively the source and target vertices of each edge $e$. A spin network state (or exactly a gauge-invariant cylindrical function of the $\SU(2)$ connection on the graph $\Gamma$) is a function $\vphi$ of $\SU(2)$ group elements $g_e$ living on each edge $e$ and satisfying a gauge invariance at each vertex:
\be
\vphi(\{g_e\})
\,=\,
\vphi(\{h_{s(e)}^{-1}g_eh_{t(e)}\}),
\quad \forall g_e\in\SU(2)^E,\quad \forall h_v\in\SU(2)^V\,.
\ee
The classical data on a graph is thus given by (the equivalence classes under $\SU(2)$ gauge transformations at the vertices of) the group elements $g_e\in\SU(2)$. It was shown that this setting can be replaced by spinors $z^v_e$ living at the vertices and labeled by the edges attached to that vertex, which allow a more direct geometrical interpretation of the classical data as defined on a discrete 3d space geometry.

\medskip

Let us start by focusing on a single intertwiner or vertex. We assume that it has $N$ edges attached to it. We attach one spinor $z_e$ to each leg of the vertex, with $e$ running from 1 to $N$. A spinor $z_e$ determines a 3-vector $\vV(z)$ through its projection on Pauli matrices:
\be
|z\ra \la z| = \f12 \left( {\la z|z\ra}\id  + \vec{V}(z)\cdot\vec{\sigma}\right),
\ee
with $|\vV(z)|=\la z| z\ra$.
Conversely the original spinor $z$ is entirely determined by the 3-vector $\vV(z)$ up to a phase. We give more details on this in Appendix, where we also define the dual spinor $|z]$ such that:
\be
[z|z\ra=0,\qquad [z|z]=\la z|z\ra,\qquad \vV(|z])=- \vV(|z\ra)\,.
\ee

The phase space is defined by simply postulating that the spinor $z_e$ is dual to its complex conjugate $\bz_e$:
\be
\{z_e^a,\bz_e^b\}=-i\delta_{ab}.
\ee
The (components of the) 3-vector $\vV(z_e)$ can be seen to generate $\SU(2)$ transformations on the spinor $z_e$ and we can check that their Poisson bracket truly form the $\su(2)$ algebra:
\be
\{V_i(z_e),V_j(z_e)\}=\,2\epsilon_{ijk}V_k(z_e)\,.
\ee
Then we further impose the closure constraints $\sum_e \vV(z_e)=0$, which generates global $\SU(2)$  transformations on the spinors $z_e$. This constraint is easily translated in terms of the spinors $z_e$:
\be
\sum_e |z_e\ra\la z_e| \,=\,\f12 \sum_e \la z_e|z_e\ra\,\id\,.
\ee
The reader can find more details in Appendix or in the references \cite{un2,un3,un4}.
Thus we can write a classical action principle which defines this constrained phase space:
\be
S_v[z_e]
\,=\,
\int dt\,
\sum_e\,\left(-i\,\la z_e|\pp_t z_e\ra + \la z_e| \Lambda| z_e\ra\right),
\ee
where $\Lambda$ is the 2$\times$2  matrix Lagrange multiplier, with $\tr\,\Lambda=0$, which imposes the closure constraint. Let us stress that this action principle only defines the kinematics on this phase space and we haven't yet added dynamics to it.

The work done in \cite{un4} was to identify suitable $\SU(2)$ observables:
\be
E_{ef}=\la z_e|z_f\ra,\qquad F_{ef}=[ z_e|z_f\ra,\qquad \bF_{ef}=\la z_f|z_e]\,.
\ee
It is clear that these quadratic combinations are invariant under global $\SU(2)$ transformations on the spinors $z_e\arr g\,z_e$ for $g\in\SU(2)$. We can also check that their Poisson bracket with the closure constraints vanish. They thus turn out to be a nice complete sets of observables on the constrained phase space. The $\U(N)$ formalism for intertwiners is actually based on promoting these observables to operators acting on the Hilbert space of intertwiner states \cite{un1,un2,un4}.

\medskip

Now, having a spinorial description of each vertex, we can glue these structures together along the edges and form a spinor network. We now have spinors $z^v_e$ around each vertex $v$, which satisfy the closure constraints around each vertex. The gluing will induce a further constraint on each edge $e$:
\be
\la z^{s(e)}_e|z^{s(e)}_e\ra\,=\,\la z^{t(e)}_e|z^{t(e)}_e\ra\,.
\ee
This ensures that the two 3-vectors living on the same edge $e$, but attached to the source and target vertices, have the same norm $|\vV(z^{s(e)}_e)|=|\vV(z^{t(e)}_e)|$ and are thus related by an $\SU(2)$ rotation, which is exactly the group element $g_e$ of the standard formulation.

The corresponding action principle, which defines the classical kinematical structure, of spinor network is \cite{un4}:
\be
S_\Gamma[z^v_e]
\,=\,
\int dt\,
\sum_{v,e}-i\la z^v_e|\pp_t z^v_e\ra + \la z^v_e| \Lambda^v| z^v_e\ra
+\sum_e \lambda^e\,\big{(}
\la z^{s(e)}_e|z^{s(e)}_e\ra\,-\,\la z^{t(e)}_e|z^{t(e)}_e\ra
\big{)}\,,
\label{actionSU2}
\ee
where the $\lambda^e$'s are Lagrangian multipliers imposing the new gluing constraints.

\subsection{The Classical Spinor Framework  for $\Spin(4)$}
\label{spin4_class}

We now adapt this spinorial framework for $\SU(2)$ to $\Spin(4)$, which is the relevant gauge group for Riemannian 4d quantum gravity. Since, $\Spin(4)\sim\SU_L(2)\times\SU_R(2)$ exactly factorizes in its left and right $SU(2)$-subgroups, we take exactly two independent copies of the spinors introduced above.

Starting around a single vertex as before, we now have two spinors for edge $e$ attached to that vertex, $z^L_e$ and $z^R_e$. The corresponding 3-vectors $\vV(z^L_e)$ and $\vV(z^R_e)$ generates respectively $\SU_L(2)$ and $\SU_R(2)$ transformations. We can combine them to reconstruct the standard $\su(2)$ generators and boost generators:
\be
\vJ_e=\f12\big{(}\vV(z^L_e)+\vV(z^R_e)\big{)},\qquad
\vK_e=\f12\big{(}\vV(z^L_e)-\vV(z^R_e)\big{)},
\ee
which satisfy the expected Poisson brackets:
\be
\{J_e^i,J_e^j\}=\epsilon^{ijk} J_e^k,\qquad
\{J_e^i,K_e^j\}=\epsilon^{ijk} K_e^k,\qquad
\{K_e^i,K_e^j\}=\epsilon^{ijk} J_e^k\,.
\ee
Therefore, an $\SU(2)$-rotation will act on the two spinors  $z^L_e$ and $z^R_e$ with the space $\SU(2)$ group element, $(z^L,z^R)\arr (g\,z^L,g\,z^R)$, while boosts will act on the two spinors with inverse group elements, $(z^L,z^R)\arr (g\,z^L,g^{-1}\,z^R)$.

Then we impose the invariance under the global $\Spin(4)$-action on the spinors around the vertex, which is generated by both left and right closure constraints $\sum_e \vV(z^L_e)=\sum_e \vV(z^R_e)=0$, which are trivially equivalent to $\sum_e \vJ_e=\sum_e \vK_e=0$.

\medskip

Now moving back to the full graph and spinor network, we have two copies of the spinors $z^v_e{}^{L,R}$ and the action principle defining the kinematical phase space structure is just the sum of the two previously defined actions \Ref{actionSU2}:
\be
\cS_\Gamma[z^v_e{}^{L,R}]
\,=\,
S_\Gamma[z^v_e{}^L]+S_\Gamma[z^v_e{}^R]\,.
\ee

\subsection{The Simplicity Constraints: Differences and Equivalence}
\label{simplicity_class}

Now we would like to discuss the simplicity constraints. They are constructed through their action on each intertwiner independently, so we will focus on one vertex $v$ and drop the index $v$ in this section. We introduce our holomorphic simplicity constraints:
\be
\forall e,f\,\quad F^L_{ef}=\rho^2 F^R_{ef}
\qquad
\textrm{i.e.}
\quad
\forall e,f\,\quad [z_e^L|z_f^L\ra=\rho^2 [z_e^R|z_f^R\ra
\ee
where $\rho$ is  a fixed parameter, related to the Immirzi parameter.
We will also refer to these as the $F$-constraints or $F$-simplicity.

A first remark is that $F_{ef}$ is anti-symmetric in $e\leftrightarrow f$, and in particular $F_{ee}$ vanishes for $e=f$. Therefore, the $F$-constraints are trivial for $e=f$ and symmetric under the exchange $e\leftrightarrow f$.

A second remark is that the $F_{ef}$ are holomorphic in the spinors, thus the name ``holomorphic simplicity constraints". In particular, their Poisson brackets with each other vanish:
\be
\big{\{}F^L_{ef}-\rho^2 F^R_{ef}\,,\,F^L_{\te\tf}-\rho^2 F^R_{\te\tf}\big{\}}\,=\,0\,.
\ee

First, we show that  $F$-simplicity implies the standard simplicity constraints:
\begin{prop}
Assuming the holomorphic simplicity constraints, $[z_e^L|z_f^L\ra=\rho^2 [z_e^R|z_f^R\ra$ for all couple of edges $e,f$, and assuming the closure constraints $\sum_e |z_e^L\ra\la z_e^L|\propto \id$ and $\sum_e |z_e^R\ra\la z_e^R|\propto \id$, then the following amongst the spinors and 3-vectors are implied:
\be
\vV_e^L\cdot\vV_f^L
\,=\,
|\rho^2|^2\,\left(\vV_e^R\cdot\vV_f^R\right),
\ee
\be
\la z_e^L|z_f^L\ra
\,=\,
|\rho^2|\,\la z_e^R|z_f^R\ra.
\ee
\end{prop}

These are the standard simplicity constraints. In particular, for $e=f$, we get the diagonal simplicity constraints, which we can express in terms of the $\spin(4)$ generators $\vJ_e$ and $\vK_e$:
\be
(1-|\rho^2|^2)\,(\vJ^2+\vK^2)+(1+|\rho^2|^2)\,(2\,\vJ\cdot\vK)
\,=\,0\,.
\ee

\begin{proof}

We are going to take the norm squared of the $F$-observables. First, one can check that:
\beq
|F_{ef}|^2
&=&
\la z_f|z_e][z_e|z_f\ra
=
\tr|z_e][z_e|z_f\ra\la z_f| \nn\\
&=&
\f14\tr \,\Big{(}|\vV_e| \id-\vV_e\cdot\vsigma\Big{)}\,\Big{(}|\vV_f| \id+\vV_f\cdot\vsigma\Big{)} \nn\\
&=&
\f12\,\Big{(}|\vV_e||\vV_f|-\vV_e\cdot\vV_f\Big{)}\,.
\eeq
Thus taking the norm squared of the $F$-simplicity constraint gives:
\be
|\vV_e^L||\vV_f^L|-\vV_e^L\cdot\vV_f^L
\,=\,
|\rho^2|^2\,\left(|\vV_e^R||\vV_f^R|-\vV_e^R\cdot\vV_f^R\right).
\ee
Summing this relation over both edges $e$ and $f$ and taking into account that the vectors satisfy the  closure condition $\sum_e \vV_e^L=\sum_e \vV_e^R=0$, we easily get:
\be
\sum_e |\vV_e^L|
\,=\,
|\rho^2|\,\sum_e |\vV_e^R|.
\ee
Then coming back to the previous relation and only summing over the index $f$ while keeping $e$ fixed, we get:
\be
\forall e,\quad |\vV_e^L| \,=\,|\rho^2|\,|\vV_e^R|.
\ee
Plugging this back in the expression of $|F_{ef}|^2$, we finally get:
\be
\forall e,f,\quad
\vV_e^L\cdot\vV_f^L
\,=\,
|\rho^2|^2\,\left(\vV_e^R\cdot\vV_f^R\right),
\ee
which are the standard quadratic simplicity conditions.

We also check that the $F$-simplicity constraints imply the $E$-simplicity constraints i.e:
\be
\forall e,f,\quad
\la z_e^L|z_\te^L\ra
\,=\,
2\,\f{\sum_f\la z_e^L|z_f^L][z_f^L|z_\te^L\ra}{\sum_f \la z_f^L|z_f^L\ra}
\,=\,
2\,\f{|\rho^2|^2\,\sum_f\la z_e^R|z_f^R][z_f^R|z_\te^R\ra}{|\rho^2|\,\sum_f \la z_f^R|z_f^R\ra}
\,=\,
|\rho^2|\,\la z_e^R|z_\te^R\ra,
\ee
where we use the fact that we already know that
$\la z_f^L|z_f^L\ra=|\vV_f^L|=|\rho^2|\,|\vV_f^R|=|\rho^2|\,\la z_f^R|z_f^R\ra$.

\end{proof}

Now, we can make the link between these new holomorphic simplicity constraints and the ``linear" simplicity constraints involving the time normal. As already explained in previous work \cite{FK,LS2,un3}, the time normal gets encoded as an $\SU(2)$ transformation between the left spinors and the right spinors.

\begin{prop}
Assuming the closure constraints, then the holomorphic simplicity constraints are equivalent to the linear simplicity constraints,  i.e. there exists a group element $g\in\SU(2)$ such that:
\be
\exists g\in\SU(2),\quad
\forall e,\quad
g\,|z_e^L\ra\,=\, \rho\,|z_e^R\ra.
\ee
\end{prop}

Let us translate this linear simplicity constraint into a constraint on the 3-vectors, which is easier to understand geometrically.
Dropping the index $e$,
the condition $g\,|z^L\ra\,=\, \rho\,|z^R\ra$ implies that $g\,|z^L\ra\la z^L|g^{-1}\,=\, |\rho|^2\,|z^R\ra\la z^R|$ i.e. that $g$ rotates the 3-vector $\vV^L$ onto its right counterpart $\vV^R$:
$$
g\,|z^L\ra\,=\, \rho\,|z^R\ra
\,\Rightarrow\,
g\vartriangleright \vV^L=|\rho|^2\,\vV^R\,.
$$
Now, let us think in term of a bivector, or Lie algebra element in $\spin(4)$. We can write a bivector $B$ either in terms of its left/right components $(\vV^L,\vV^R)$ or in terms of its rotation/boost components $(\vJ,\vK)$, with the correspondence given by $\vV^L=\vJ+\vK$ and $\vV^R=\vJ-\vK$. The Hodge dual of the bivector $\star B$ is obtained by switching the rotation and boost components $(\vK,\vJ)$ or by simply switching the sign of its right part $(\vV^L,-\vV^R)$.

We consider the combination $B+\gamma\star B=((1+\gamma)\vV^L,(1-\gamma)\vV^R)$. The parameter $\gamma$ is the Immirzi parameter. We act on $B+\gamma\star B$ with the $\Spin(4)$ transformation $G=g_{(L)}\otimes \id_{(R)}$:
$$
G\vartriangleright (B+\gamma\star B)
\,=\,
(G\vartriangleright B)+\gamma\star (G\vartriangleright B)
\,=\,
((1+\gamma)\,g\vartriangleright \vV^L\,,\,(1-\gamma)\,\vV^R).
$$
We distinguish two cases:
\begin{itemize}

\item {\bf $|\gamma|<1$}: Then we take $|\rho|^2=(1-\gamma)/(1+\gamma)\,>0$. Then the boost part of $G\vartriangleright (B+\gamma\star B)$ vanishes (i.e its left component is equal to its right component). Thus the linear simplicity constraint $g\,|z_e^L\ra\,=\, \rho\,|z_e^R\ra$ for all $e$ means that there exists a common time normal to all bivectors:
    \be
    \forall e,\quad\cN^I\,(B_e+\gamma\star B_e)_{IJ}=0,
    \qquad
    \cN=G^{-1}\vartriangleright (1,0,0,0)\,.
    \ee

\item {\bf $|\gamma|>1$}: Because of the sign switch, we take $|\rho|^2=(\gamma-1)/(1+\gamma)\,>0$. Then the rotation part of $G\vartriangleright (B+\gamma\star B)$ vanishes (i.e its left component is equal to minus its right component), or equivalently the boost part of $G\vartriangleright (\star B+\gamma B)$ vanishes. Thus the linear simplicity constraint $g\,|z_e^L\ra\,=\, \rho\,|z_e^R\ra$ for all $e$ means once again that there exists a common time normal to all bivectors:
    \be
    \forall e,\quad\cN^I\,( B_e+\f1\gamma \star B_e)_{IJ}=0,
    \qquad
    \cN=G^{-1}\vartriangleright (1,0,0,0)\,.
    \ee

\end{itemize}

Let us now get back to proving the previous proposition.

\begin{proof}

To start with, it is easy to see that the existence of such a group element implies $F$-simplicity since the $F_{ef}$'s are $\SU(2)$-invariant observables:
\be
[z_e^L|z_f^L\ra=\rho^2 \,[z_e^R|g\,g^{-1}|z_f^R\ra=\rho^2 \,[z_e^R|z_f^R\ra\,.
\ee
It is actually straightforward to show the converse statement. It means that the $F$-observables are a complete set of $\SU(2)$-invariant observables.

Let's start by assuming $F$-simplicity. Let's choose one index $e$ and consider the two spinors $z_e^{L,R}$. The $F$-simplicity implies that the ratio of the norms of these two spinors is given by $|\rho^2|$, $\la z_e^L|z_e^L\ra=|\rho^2|\,\la z_e^R|z_e^R\ra$. Then there exists a (unique) $\SU(2)$ group element which maps one onto the other (the interested reader can find more details in Appendix):
\be
g_e\,\equiv\,\f{|z_e^R\ra\la z_e^L|+|z_e^R][z_e^L|}{\sqrt{\la z_e^L|z_e^L\ra\,\la z_e^R|z_e^R\ra}},\qquad
g_e|z_e^L\ra\,=\, |\rho|\,|z_e^R\ra.
\ee
For the sake of simplicity, we will now assume that $\rho\in\R^+$ is real and (strictly) positive \footnotemark. Then we can check that this group element $g_e$ actually maps any spinor $z_f^L$ to its right counterpart $z_f^R$~:
\be
g_e\,|z_f^L\ra
\,=\,
\f{|z_e^R\ra\la z_e^L|z_f^L\ra+|z_e^R][z_e^L|z_f^L\ra}{\rho\,\la z_e^R|z_e^R\ra}
\,=\,
\rho^2\,\f{|z_e^R\ra\la z_e^R|z_f^R\ra+|z_e^R][z_e^R|z_f^R\ra}{\rho\,\la z_e^R|z_e^R\ra}
\,=\,
\rho\,|z_f^R\ra,
\ee
since $|z_e^R\ra\la z_e^R|+|z_e^R][z_e^R|=\la z_e^R|z_e^R\ra\,\id$.
\footnotetext{
We can similarly treat the generic case of a complex parameter $\rho$ by being careful with phase factors. Writing $\rho=r\,e^{i\theta}$, $F$-simplicity implies that $\la z_e^L|z_f^L\ra=\,r^2\,\la z_e^R|z_f^R\ra$. Then we define the group element $g_e$ such that $g_e|z_e^L\ra\,=\, \rho\,|z_e^R\ra\,=\,r\,e^{i\theta}\,|z_e^R\ra$:
$$
g_e\,\equiv\,\f{e^{i\theta}|z_e^R\ra\la z_e^L|+e^{-i\theta}|z_e^R][z_e^L|}{r\,\la z_e^R|z_e^R\ra}.
$$
Then we can check its action on all other left spinors:
$$
g_e\,|z_f^L\ra
\,=\,
\f{e^{i\theta}|z_e^R\ra\la z_e^L|z_f^L\ra+e^{-i\theta}|z_e^R][z_e^L|z_f^L\ra}{r\,\la z_e^R|z_e^R\ra}
\,=\,
\f{r^2e^{i\theta}|z_e^R\ra\la z_e^R|z_f^R\ra +e^{-i\theta}r^2e^{2i\theta}|z_e^R][z_e^R|z_f^R\ra}{r\,\la z_e^R|z_e^R\ra}
\,=\,
\rho\,|z_f^R\ra.
$$
}

\end{proof}

This shows the equivalence between the $F$-simplicity and the linear simplicity constraints, but also provides us with the explicit expression for the time normal, or equivalently the group element mapping left spinors to right spinors, in terms of those spinors.

This construction also allows us to see how to go between the two sectors that we distinguished above with $|\gamma|<1$ and $|\gamma|>1$. Indeed, if we now assume that we have a group element such $g\,|z_e^L\ra\,=\, \rho\,|z_e^R]$ for all $e$'s, then it is equivalent to requiring the conjugate $F$-simplicity $F^L_{ef}=\rho^2 \overline{F^R_{ef}}$. This condition implies that $g\vartriangleright \vV^L=-|\rho|^2\,\vV^R$. This sign switch allows us to swap the two cases $|\gamma|<1$ and $|\gamma|>1$.

%
%
%
%
%

\subsection{Classical Phase Space for Simple Intertwiners}

Let us look at the classical $\Spin(4)$-invariant phase space for a single intertwiner and impose the holomorphic simplicity constraints. Thus we start with the (free) action:
\be
S_v^{\textrm{simple}}[z_e^{L,R}]
\,=\,
\int dt\,
\sum_e\sum_{\eta=L,R}\,-i\,\la z_e^\eta|\pp_t z_e^\eta\ra +
\la z_e^\eta| \Lambda^\eta| z_e^\eta\ra
+\sum_{e,f} \phi_{ef}\,\left([z_e^L|z_f^L\ra-\rho^2\,[z_e^R|z_f^R\ra \right)\,,
\ee
where $\phi_{ef}$ is the Lagrange multiplier imposing the simplicity constraints.
Then as was shown it above, we can solve these constraints exactly and write the right spinors in terms of the left spinors:
$$
\exists g\in\SU(2),\quad
\forall e,\quad
|z_e^R\ra \,=\, \rho^{-2}\,g\,|z_e^L\ra.
$$
This group element $g$ is then re-absorbed in the Lagrange multiplier $\Lambda^R$ and we are left with solely the left sector:
\be
S_v^{\textrm{simple}}[z_e^L]
\,=\,
\int dt\,
\sum_e\,-2i\,\la z_e^L|\pp_t z_e^L\ra +
\la z_e^L| \Lambda^L| z_e^L\ra\,.
\ee
This reduces to the phase space for a single $\SU(2)$ intertwiner.
This shows that imposing the holomorphic simplicity constraints on a single intertwiner reduces effectively the degrees of freedom down back to the $\SU(2)$ theory.

We can further move to a full spinor network on a graph $\Gamma$. A priori, we have to be careful with the gluing condition. We start by solving for the right spinors in terms of the left spinors. This gives us an action which depends on the left spinors $z_e^v{}^L$ and on group elements $g_v$ at each vertex. Then we realize that the gluing conditions involve only norms of the spinors and thus do not see the group elements $g_v$ at all. And finally, we arrive back at the action for a pure $\SU(2)$ spinor network.
%

This shows that, at the kinematical level, the holomorphic simplicity constraints allow to reduce effectively the $\Spin(4)$ phase space down to the $\SU(2)$ phase space. This means that, after quantization, at the kinematical level, the  simple $\Spin(4)$ spin network states should be in one-to-one correspondence with $\SU(2)$ spin network states. This is actually a desired feature from the point of view that spinfoam models based on simple $\Spin(4)$ spin networks should compute transition amplitudes for the $\SU(2)$ spin networks of loop quantum gravity \cite{EPR,EPRL}.

We would like to underline that this is at the kinematical level and that we expect that the dynamics of the theory will reflect the $\Spin(4)$  structure and invariance of the theory.

\section{Coherent States and Simplicity at the Quantum Level}

In this section, we will quantize all the classical spinorial structures defined in the previous section. This will lead us to $\SU(2)$ and $\Spin(4)$ intertwiner states and to the quantum simplicity constraints. We will then show how to solve these holomorphic simplicity constraints using coherent intertwiner states, which we will dub simple intertwiners.

Our starting point is the work on the $\U(N)$ formalism for $\SU(2)$ intertwiners \cite{un1,un2} and on solving the quantum simplicity constraint using $\U(N)$ coherent states \cite{un3}. We will review these previous results in a concise and consistent fashion. We will also describe how to glue coherent intertwiners into coherent $\SU(2)$ spin network states and how to similarly glue simple intertwiners into simple spin networks. This will set the proper foundations in order to build the corresponding spinfoam amplitudes, as we will do in the next section.

\subsection{$\SU(2)$-Intertwiner Spaces and $\U(N)$ Representations}
\label{irreps}

Let us start with the quantization of the classical phase space for $\SU(2)$.
We will focus on a single vertex $v$ of valence $N$ and we will drop the index $v$. This is a review of the $\U(N)$ formalism developed in \cite{un0,un1,un2}.

Quantizing the spinor phase space can be done in a straightforward way. Considering a spinor $z$ and its canonically conjugated $\bz$, we quantize its two components as creation/annihilation operators of two harmonic oscillators:
\be
\begin{array}{c}
z^0\arr a\\
z^1\arr b
\end{array}
\qquad
\begin{array}{c}
\bz^0\arr a^\dag\\
\bz^1\arr b^\dag
\end{array}
\qquad\quad
[a,a^\dag]=[b,b^\dag]=1,\quad[a,b]=0\,.
\ee
So, on each leg $e$ around the vertex $v$, we have a couple of harmonic oscillators $a_e,b_e$, and we use the standard basis $|n^a_e,n^b_e\ra$ labeled with the number of quanta for both oscillators.

Using this quantization procedure, we directly quantize the vectors $\vV(z_e)$ and the observables $E_{ef},F_{ef}$. The components of the 3-vectors are given by the Schwinger  representation of the $\su(2)$ algebra in term of a couple of harmonic oscillators:
\be
\begin{array}{lcl}
\f12V^3=\f12\left(|z^0|^2-|z^1|^2\right)
&\,\arr\,&
\f12\hat{V}^3= \f12\left(a^\dag a - b^\dag b\right) \\
\f12V^+=\bz^0z^1
&\,\arr\,&
\f12\hat{V}_+= a^\dag  b \\
\f12V^-=z^0 \bz^1
&\,\arr\,&
\f12\hat{V}^-= a b^\dag \\
|V|=\la z|z\ra=\left(|z^0|^2+|z^1|^2\right)
&\,\arr\,&
\widehat{|V|}= \left(a^\dag a + b^\dag b\right)
\end{array}
\ee
The components of $\vV(z_e)$ form a  $\su(2)$ Lie algebra as expected from the Poisson brackets.
So we now have each leg $e$ carries an $\SU(2)$-representation.

The norm operator $\widehat{|V(z_e)|}$ is the total energy of the oscillators and is $\SU(2)$-invariant. Fixing its value projects us onto an irreducible $\SU(2)$-representation and its value gives twice the spin $j_e$ of that representation.
More precisely, we go from the standard oscillator basis $|n^a_e,n^b_e\ra$ to the usual magnetic momentum basis $|j_e,m_e\ra$ for spin systems by diagonalizing the operators $\widehat{V^3_e}$ and $\widehat{|V_e|}$, which gives the simple correspondence:
\be
j_e=\f{n^a_e+n^b_e}{2},
\qquad
m_e=\f{n^a_e-n^b_e}{2}.
\ee
So fixing the total energy of the two harmonic oscillators, we fix the spin $j_e$ of the $\SU(2)$-representation attached to the leg $e$.
Calling $\cH^{HO}=\oplus_{n}\C\,|n\ra$ the Hilbert space of a single harmonic oscillator, we can write more generally:
\be
\cH^{HO}\otimes\cH^{HO}=\bigoplus_{j\in\N/2}\cV^j,
\ee
where we write $\cV^j$ for the $\SU(2)$-representation of spin $j$.

The next step is to impose the closure constraints $\sum_e \widehat{\vV}_e=0$. This means imposing the invariance under the global $\SU(2)$-action, which implies considering $\SU(2)$-invariant states in the tensor product of the $\SU(2)$-representations living on the legs $e$ around the vertex, i.e intertwiners between the spins $j_e$. This leads to the whole Hilbert space of $N$-valent intertwiners:
\be
\label{int_space}
\cH_N
\,=\,
\textrm{Inv}_{\SU(2)}\bigotimes_e^N (\cH^{HO}_e\otimes\cH^{HO}_e)
\,=\,
\textrm{Inv}_{\SU(2)}\bigotimes_e \bigoplus_{j_e\in\N/2}\cV^{j_e}
\,=\,
\bigoplus_{\{j_e\}} \textrm{Inv}_{\SU(2)} \bigotimes_e \cV^{j_e}.
\ee

Then we quantize the observables $E_{ef},F_{ef}$ following the same quantization procedure:
\be
\begin{array}{lcl}
E_{ef}=\la z_e|z_f\ra
&\,\arr\,&
\hE_{ef}=a^\dag_e a_f+ b^\dag_e b_f\\
F_{ef}=[ z_e|z_f\ra
&\,\arr\,&
\hF_{ef}=a_e b_f- b_e a_f \\
\bar{F}_{ef}=\la z_f|z_e]=-\la z_e|z_f]
&\,\arr\,&
\hFd_{ef}= a^\dag_e b^\dag_f- b^\dag_e a^\dag_f
\end{array}
\ee
It is straightforward to compute the commutators between these observables and to check that their algebra closes \cite{un2}. Moreover, as was shown in \cite{un4}, these commutators provide the expected quantization of their classical Poisson brackets. The interested reader can find the relevant Poisson brackets and commutators in appendix \ref{app_commEF}.

It is also straightforward to check that these operators $\hE_{ef},\hF_{ef},\hFd_{ef}$ commute  with the $\SU(2)$ generators $\sum_e \widehat{\vV}_e$, so that they are still $\SU(2)$-invariant observables at the quantum level.

\medskip

The important fact is that the $\hE$-operators form a $\u(N)$ Lie algebra. This comes from the fact that the $E_{ef}$ generate $\U(N)$-transformations on the spinors trough the Poisson bracket at the classical level (as was checked in \cite{un2,un3,un4}):
\be
\{E_{ef},z_l\}
\,=\,
i\delta_{el}z_f\,,
\qquad
e^{\sum_{e,f}\alpha_{ef}\{E_{ef},\cdot\}}\vartriangleright z_l
=\sum_f (e^{i\alpha})_{lf}z_f=(e^{i\alpha}z)_l,
\ee
where $\alpha=\alpha^\dag$ is an $N\times N$ Hermitian matrix and $U=e^{i\alpha}$ the corresponding unitary transformation in $\U(N)$. Similarly, at the quantum level, the operators $\hE_{ef}$ generate a $\U(N)$-action on the space of $N$-valent intertwiners \cite{un0,un1}. Without going into the details (which the interested reader will find in \cite{un1}), the final result is that each space of intertwiners at fixed total area $J=\sum_e j_e$ carries an irreducible representation of $\U(N)$ (whose highest weight vector is a bivalent intertwiner). This is summarized by the following decomposition of the intertwiner space:
\be
\cH_N
\,=\,
\bigoplus_{\{j_e\in\N/2\}} \textrm{Inv}_{\SU(2)} \bigotimes_e \cV^{j_e}
\,=\,
\bigoplus_{J\in\N} \cR^J\,,\qquad
\cR^J
\,=\,
\bigoplus_{J=\sum_ej_e} \textrm{Inv}_{\SU(2)} \bigotimes_e \cV^{j_e}\,.
\ee
The $\u(N)$-generators $\hE_{ef}$ act within each $\U(N)$-representation $\cR^J$ at fixed $J$, while the operators $\hF_{ef}$ (resp. $\hFd_{ef}$) act as annihilation operators (resp. creation operators) and allow transitions from a subspace $\cR^J$ to $\cR^{J-1}$ (resp. $\cR^{J+1}$).

\subsection{Coherent Intertwiner States}



Following the logic of the series of papers \cite{un1,un2,un3,un4} on the $\U(N)$ formalism for $\SU(2)$ intertwiners, we introduce coherent intertwiner states which are peaked on each point of the spinor phase space. We start with $\SU(2)$ coherent states and review the various definitions of coherent intertwiners until the most advanced one which will allow to solve the holomorphic simplicity constraints.

We start by introducing $\SU(2)$ coherent states living in each $\SU(2)$-representation at fixed spin $j$. We define them by acting with the relevant creation operators on the vacuum of the harmonic oscillators:
\be
|j,z\ra
\,=\,
\f{(z^0 a^\dag+z^1 b^\dag)^{2j}}{\sqrt{(2j)!}}\,|0\ra
\,=\,
\sum_{m=-j}^{+j}\f{\sqrt{(2j)!}}{\sqrt{(j+m)!(j-m)!}}\,(z^0)^{j+m}(z^1)^{j-m}\,|j,m\ra\,.
\ee
It is pretty easy to compute the norm of these vectors:
\be
\la j,z|j,z\ra
\,=\,
\la z|z\ra^{2j}\,.
\ee
These are coherent states under the $\SU(2)$-action, i.e they transform covariantly under $\SU(2)$-transformations (see e.g. \cite{un2,un3}):
\be
\forall g\in\SU(2),\quad
g\,|j,z\ra
=|j,g\,z\ra\,,
\ee
where the action of $g\in\SU(2)$ on the spinor $z$ is simply the standard action of the $2\times 2$ matrix (in the fundamental representation). From this, we can simply deduce another fundamental property of these $\SU(2)$ coherent states: they can all be obtained from the highest weight vector $|j,j\ra$ by acting with $\SU(2)$ group element. Indeed, an arbitrary spinor $z$ can always be obtained from the ``origin spinor" $\Omega=(1,0)$ by a (unique) $\SU(2)$ transformation:
\be
g(z)\,|\Omega\ra=\f{|z\ra}{\sqrt{\la z|z\ra}},\qquad
g(z)=\f1{\sqrt{\la z|z\ra}}\,\mat{cc}{z^0 & -\bz^1\\ z^1 &\bz^0}
=\f1{\sqrt{\la z|z\ra}}\,(|z\ra,|z])
\ee
This implies a similar relation on the coherent states:
\be
\f1{\sqrt{\la z|z\ra^{2j}}}|j,z\ra=g(z)\,|j,\Omega\ra=g(z)\,|j,j\ra.
\ee
From here, taking into account that $|j,j\ra=|\f12,\f12\ra^{\otimes 2j}$, we deduce the tensoring properties of the $\SU(2)$ coherent states:
\be
|j,z\ra=|\f12,z\ra^{\otimes 2j},\qquad
|\f12,z\ra=|z\ra\,.
\ee
Finally, these states are semi-classical and the expectation value of the $\su(2)$-generators $\vJ$ on them is as expected:
\be
\f{\la j,z|\vJ|j,z\ra}{\la j,z|j,z\ra}
\,=\,
2j\,\f{\la z|\f{\vsigma}{2}|z\ra}{\la z|z\ra}
\,=\,
j\,\f{\vV(z)}{|\vV(z)|}\,.
\ee

\medskip

The first notion of coherent intertwiner was then introduced in \cite{LS} from tensoring together $\SU(2)$ coherent states and group-averaging in order to get $\SU(2)$-invariant states. This was re-cast in terms of spinors in \cite{un2,un3}. Thus such an $N$-valent coherent intertwiner is labeled by a list of $N$ spins $j_e$ and $N$ spinors $z_e$ attached to each leg $e$ and we define:
\be
|\{j_e,z_e\}\ra=
\bigotimes_e |j_e,z_e\ra,
\qquad
||\{j_e,z_e\}\ra=
\int_{\SU(2)}dg\, g\vartriangleright\bigotimes_e |j_e,z_e\ra\,.
\ee
As shown in \cite{LS}, these states have nice semi-classical properties. They are used to construct semi-classical spin network states and to define the EPRL-FK spinfoam models \cite{FK,LS2}. Nevertheless, most of their peakedness properties are only known approximatively in the large spin asymptotic regime $j_e \gg 1$ (through saddle point approximations \cite{LS}).

\medskip

Then the reference \cite{un2} introduced a new class of coherent intertwiners whose properties are under much better control than the previous coherent intertwiners constructed through group-averaging. These coherent intertwiners are covariant under the $\U(N)$-action and are constructed from the vacuum state using the $\hFd$ as creation operators. These $\U(N)$ coherent intertwiner states are labeled by their total area $J\in\N$ and $N$ spinors $z_e$ living on each leg $e$ of the intertwiners. They are defined as:
\be
|J,\{z_e\}\ra
\,=\,
\f{1}{\sqrt{J!(J+1)!}}\,
\left(\f12\sum_{e,f}[z_e|z_f\ra\,\hFd_{ef}\right)^J\,|0\ra\,.
\ee
Then it is possible to show that they are given by a superposition of the previous coherent intertwiners defined by Livine-Speziale (LS) as proved in \cite{un2}:
\be
\f{1}{\sqrt{J!(J+1)!}}\,|J,\{z_e\}\ra
\,=\,
\sum_{J=\sum_e j_e}\f{1}{\sqrt{\prod_e (2j_e)!}}\,||\{j_e,z_e\}\ra\,.
\ee
Due to their definition in terms of the creation operators $\hFd$, it is straightforward to prove that these states are covariant under the $\U(N)$-action \cite{un2}:
\be
\label{UNtransf1}
\hat{U}\,|J,\{z_e\}\ra
\,=\,
|J,\{(Uz)_e\}\ra,\qquad
U=e^{i\alpha},
\quad
\hat{U}=e^{i\sum_{e,f}\alpha_{ef}\hE_{ef}}\,,
\ee
where $\alpha$ is an arbitrary  $N\times N$ Hermitian matrix.
The behavior of these states under global rescaling of the spinors is also very simple:
\be
|J,\{\beta\,z_e\}\ra\,=\,
\beta^{2J}\,|J,\{z_e\}\ra, \quad \forall \beta \in \C\,.
\ee
We can also compute explicitly their scalar products and norms:
\beq
\la J,\{w_e\}|J,\{z_e\}\ra
&=&
\det\,\left(\sum_e |z_e\ra\la w_e|\right)^J
=
\left(\f12\sum_{e,f} \la z_f|z_e][ w_e|w_f\ra\right)^J,\nn\\
\la J,\{z_e\}|J,\{z_e\}\ra
&=&
\left(\f12\sum_e \la z_e|z_e\ra\right)^{2J}
=A(z)^{2J}\,,
\eeq
where we have introduced the new notation $A(z)$ for the sake of shortening the equations:
$$
A(z)=\f12\sum_e\la z_e|z_e\ra=\f12\sum_e |\vV(z_e)|\,.
$$
Finally, the property which will be most interesting to us in the following is that the action of the annihilation operators $\hF_{ef}$ on those states can be simply computed \cite{un2,un3,un4} (using either of two definitions of the coherent states in terms of the creation operators $\hFd$ or in terms of the LS coherent intertwiners):
\be
\hF_{ef}\,|J,\{z_e\}\ra
\,=\,
\sqrt{J(J+1)}\,[z_e|z_f\ra\,|(J-1),\{z_e\}\ra\,.
\ee
All these properties allow the exact calculation of the expectation values of the $\hE$ operators on these new coherent states \cite{un2}:
\be
\f{\la J,\{z_e\}|\hE_{ef}\,|J,\{z_e\}\ra}{\la J,\{z_e\}|J,\{z_e\}\ra}=
J\,\f{\la z_e |z_f\ra}{A(z)}.
\ee
Let us stress that this expectation value is exact while the expectation values of the $\SU(2)$-observables on the LS coherent intertwiners are only known approximatively in the large spin asymptotic limit.

\medskip

The ultimate notion of coherent intertwiners that we would like to introduce are the ones that were found in \cite{un3} to be useful when investigating the simplicity constraints at the quantum level. They are simply defined as the eigenstates of the annihilation operators $\hF_{ef}$, just as in the case of the harmonic oscillator. This is possible since the operators $\hF_{ef}$ all commute with each other. More precisely, we define following \cite{un3} coherent states labelled by a complex number $\lambda$ and the set of $N$ spinors $\{z_e\}$:
\beq
\label{coh_decomp}
|\lambda,\{z_e\}\ra
&=&
\sum_J\f{\lambda^{2J}}{\sqrt{J!(J+1)!}}\,|J,\{z_e\}\ra\\
&=&
\sum_{\{j_e\}}\prod_e\f{\lambda^{2j_e}}{\sqrt{(2j_e)!}}\,||\{j_e,z_e\}\ra\nn\\
&=&
\int dg\,g\vartriangleright
e^{\lambda\,\sum_ez_e^0 a_e^\dag+z_e^1 b_e^\dag}
\,|0\ra\nn,
\eeq
where we have used the explicit expression of the LS coherent intertwiners as the group averaging of the tensor product of $\SU(2)$ coherent states. The last equality shows the clear relation between our coherent states $|\lambda,\{z_e\}\ra$ and the standard (unnormalized) coherent states for the harmonic oscillators.

Using the previous action of the annihilation operators on the $|J,\{z_e\}\ra$ states \footnotemark, we easily check that:
\be
\hF_{ef}\,|\lambda,\{z_e\}\ra
\,=\,
\lambda^2\,[z_e|z_f\ra\,|\lambda,\{z_e\}\ra
\,=\,
\lambda^2\,F_{ef}\,|\lambda,\{z_e\}\ra\,.
\ee
\footnotetext{
Since the operator $\hF_{ef}$ is $\SU(2)$-invariant and thus commute with the $\SU(2)$-action, we could more simply compute its commutator with the usual operator $e^{\lambda\,\sum_ez_e^0 a_e^\dag+z_e^1 b_e^\dag}$. Since we have:
$$
\Big{[}\hF_{ef},\sum_lz_l^0 a_l^\dag+z_l^1 b_l^\dag\Big{]}=(z_e^0b_f+z^1_fa_e-z^0_fb_e-z^1_ea_f),
\qquad
\Big{[}(z_e^0b_f+z^1_fa_e-z^0_fb_e-z^1_ea_f),\sum_lz_l^0 a_l^\dag+z_l^1 b_l^\dag\Big{]}=2F_{ef}\,\id\,.
$$
this leads back to the same result.
}
At this point, we notice that we have $|\beta\lambda,\{z_e\}\ra=|\lambda,\{\beta z_e\}\ra$ due to the trivial scaling of these states. Therefore, we choose to set $\lambda=1$  in  the definition of $|\lambda,\{z_e\}\ra$ without loss of generality; that is, we work in the following with the states:
\be
||\{z_e\}\ra\,\equiv\,|\lambda=1\,,\{z_e\}\ra \,,
\ee
which only refer to the $N$ spinors. 

We can compute the norm and scalar product of these states\footnotemark \cite{un3}:
\beq
\la\{z_e\}||\{z_e\}\ra&=&\sum_J\f{A(z)^{2J}}{J!(J+1)!}
=\f{I_1(2A(z))}{A(z)} \\
\la\{w_e\}||\{z_e\}\ra&=&\sum_J\f{1}{J!(J+1)!}\,
\la J,\{w_e\}|J,\{z_e\}\ra
=\sum_J\f{1}{J!(J+1)!}\,
\left(\det \sum_e |z_e\ra\la w_e|\right)^J
\eeq
\footnotetext{
An interesting particular case is when the two sets of spinors only differ through a constant phase $e^{i\phi}$:
$$
\la \lambda,\{z_e\}|\lambda,\{e^{i\phi}z_e\}\ra=\sum_J\f{(e^{i\phi}|\lambda^2|A(z))^{2J}}{J!(J+1)!}
=e^{-i\phi}\f{I_1(2e^{i\phi}|\lambda|^2A(z))}{|\lambda|^2A(z)}.
$$
}
and also the expectation value of the $\hE$-operators:
\beq
\label{Evalue}
\la\{z_e\}||\hE_{ef}\,||\{z_e\}\ra
&=&
\f{\la z_e |z_f\ra}{A(z)}\sum_{J\ge 1} \f{(A(z))^{2J}}{(J-1)!(J+1)!}
\,=\,
\f{\la z_e |z_f\ra}{A(z)}\,I_2(2A(z)), \\
\f{\la\{z_e\}||\hE_{ef}\,||\{z_e\}\ra}{\la \{z_e\}||\{z_e\}\ra}
&=&
\f{\la z_e |z_f\ra}{A(z)}\,\f{A(z)\,I_2(2A(z))}{I_1(2A(z))}\,,\nn
\eeq
where $I_n$ is the modified Bessel function of the first kind. Once again, we would like to underline the fact that these expectation values are computed exactly, while the expectation values of the LS coherent intertwiners (used as a basis to build the EPRL-FK spinfoam models) are only computed (up to now) approximatively at leading order in the large spin asymptotic limit.

\begin{figure}[h]
\begin{center}
\includegraphics[height=30mm]{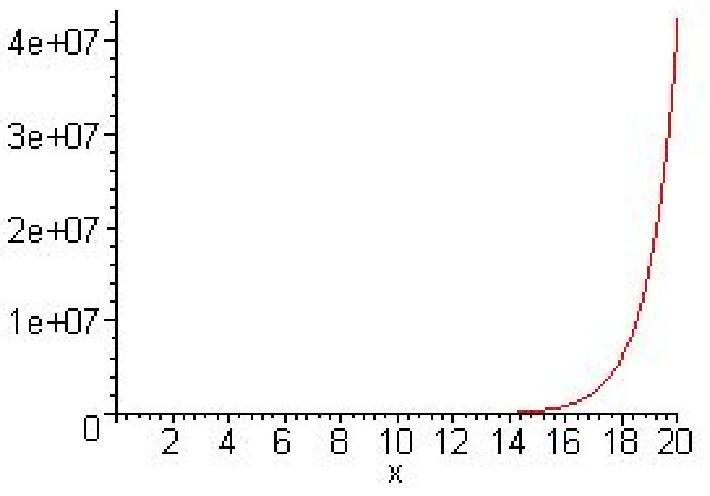}
\hspace*{5mm}
\includegraphics[height=30mm]{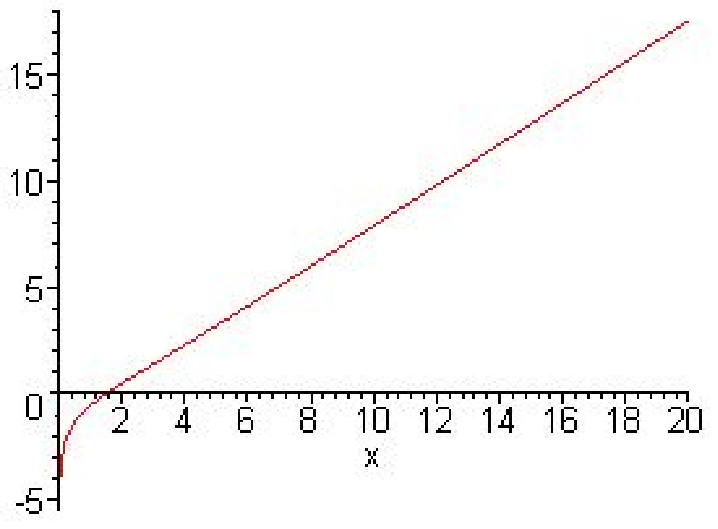}
\hspace*{5mm}
\includegraphics[height=30mm]{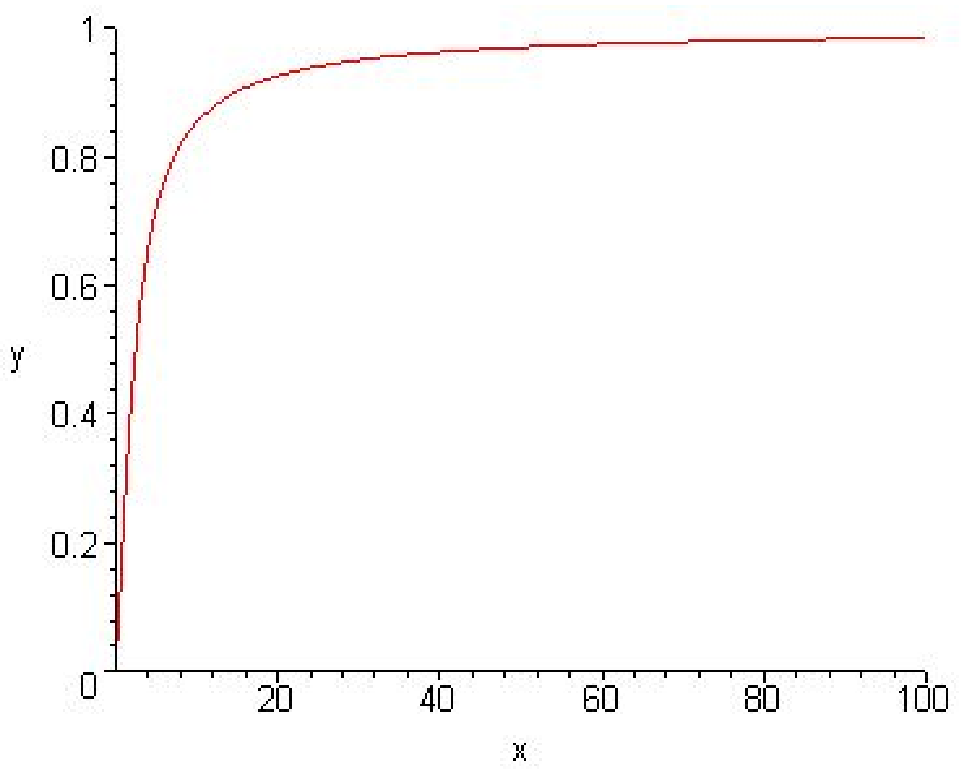}
\caption{From left ro right, plots of the modified Bessel function $I_1(x)$, of its logarithm $\ln I_1(x)$ which illustrates its asymptotic behavior for large $x\gg 1$, and the ratio $I_2(x)/I_1(x)$ which quickly converges to $1$ as $x\arr+\infty$.\label{Besselplot}}
\end{center}
\end{figure}

In our case, the asymptotic behavior of the $|| \{z_e\}\ra$ coherent states is given by
\beq
\la \{z_e\}||\{z_e\}\ra
&\sim&
\f{e^{2A(z)}}{\sqrt{4\pi}\,A(z)^{3/2}} \, ,
 \\
\f{\la\{z_e\}||\hE_{ef}\,||\{z_e\}\ra}{\la \{z_e\}||\{z_e\}\ra}
&\sim&
{\la z_e |z_f\ra}
\,,
\eeq
where the asymptotics\footnotemark{} are taken for large area $A(z)\gg 1$ (see fig.\ref{Besselplot} for plots of the Bessel functions).
\footnotetext{
The asymptotic for the modified Bessel functions $I_n(x)$ for $x\in\R$ do not depend on the label $n$ at leading order:
$$
I_n(x)\,\underset{x\arr+\infty}{\sim}\,\f{e^x}{\sqrt{2\pi x}}\,\left(1+\cO\left(\f1x\right)\right).
$$
}

\begin{figure}[h]
\begin{center}
\includegraphics[height=35mm]{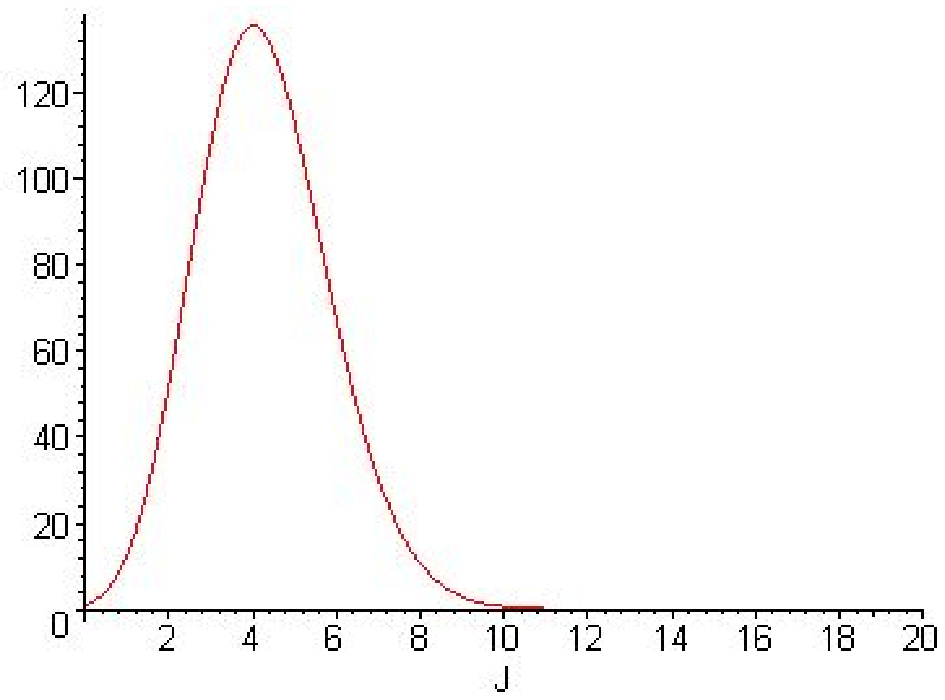}
\hspace*{5mm}
\includegraphics[height=35mm]{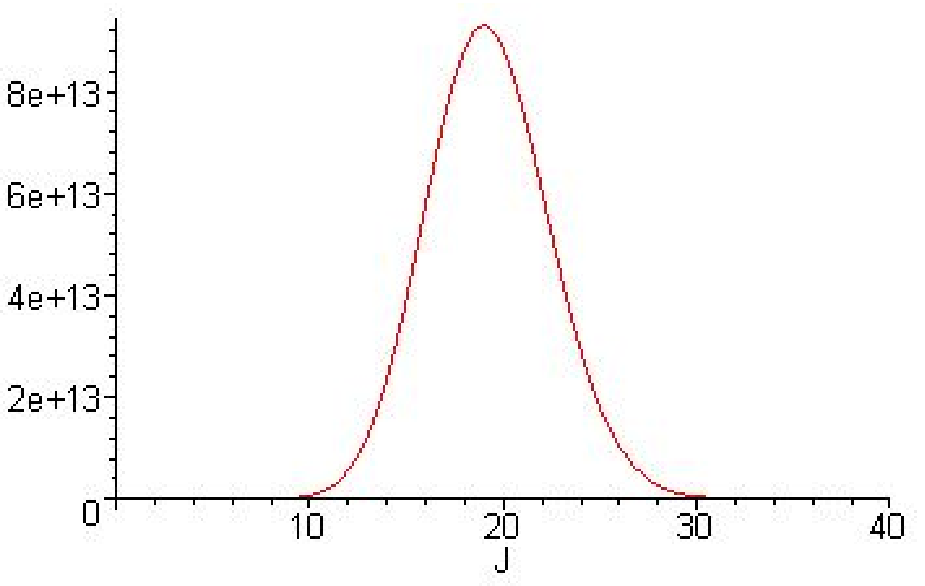}
\hspace*{5mm}
\includegraphics[height=35mm]{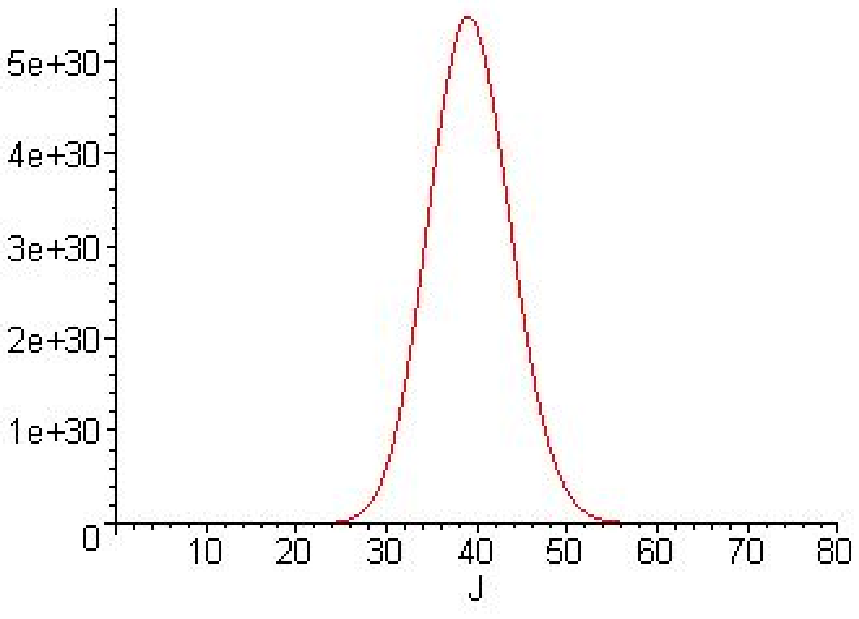}
\caption{Plots of the (un-normalized) probability distribution $P_A[J]$ for the area $J$ for various values of $A(z)$: from left to right $A(z)$ is 5, 20 and 40. We see that these distributions are approximatively Gaussians centered around the classical value $A(z)$.\label{distriJ}}
\end{center}
\end{figure}

To better understand these expectation values, let us have a look at the probability distributions on the area $J$ and spin labels $j_e$ induced by these coherent intertwiners $||\{z_e\}\ra$. We start with the area $J$, which is given as an operator by $\hat{J} \, \equiv \,\f12\sum_f E_{ff}$. Copying the formulas above, we have:
\be
\f{\la\{z_e\}||\hat{J}||\{z_e\}\ra}{\la \{z_e\}||\{z_e\}\ra}
=
A(z)\,\f{\,I_2(2A(z))}{I_1(2A(z))},
\qquad
\la\{z_e\}||\hat{J}||\{z_e\}\ra
=
\sum_JJ\,\f{A(z)^{2J}}{J!(J+1)!}\,.
\ee
%
The (un-normalized) probability distribution for the area observable $J$ is thus:
\be
P_A[J]=\f{A(z)^{2J}}{J!(J+1)!}.
\ee
We can approximate it at large $J$ by using Stirling's formula for the factorials:
$$
P_A[J]
\sim
\f1{2\pi J^2}\left(\f{eA(z)}{J}\right)^{2J}
\sim
\f1{2\pi} e^{2J(\ln A(z)+1)-(2J+2)\ln J}\,.
$$
Calling $S[J]\,\equiv\,2J(\ln A(z)+1)-(2J+2)\ln J$  the exponent, we can investigate its behavior and check whether it has any extremum:
$$
\pp_J S =2\left(\ln A(z)-\ln J-\f{1}{J} \right),
\qquad
\pp^2_JS=-\f2J\left(1-\f{1}{J}\right)\,.
$$
Thus S has a unique extremum, which is a maximum, for the value $J_0$, which is approximately $J_0\sim A(z)$ when $A(z)\gg1$. Thus for a large classical area $A(z)$, we do recover that the probability distribution for $J$ is peaked on this classical value $J_0\sim|\lambda|^2$. Moreover, we have approximatively a Gaussian around this value:
\be
P_A[J]\sim e^{S[J_0]}\,e^{-\f1{A(z)}(J-A(z))^2}\,.
\ee
We can directly see this very simple behavior on numerical simulations of this probability distribution on fig.\ref{distriJ}.

We can do a similar analysis for the individual spin labels $j_f$ associated to each leg $f$ of the intertwiners. The corresponding operator is $\hat{j_f}\,\equiv\,\f12 E_{ff}$ and we can compute:
\be
\f{\la \{z_e\}||\hat{j_f}||\{z_e\}\ra}{\la \{z_e\}||\{z_e\}\ra}
=
\f12\,{\la z_f|z_f\ra}\,\f{\,I_2(2A(z))}{I_1(2A(z))}.
\ee
Going step by step, we can derive the probability distribution on the spin label $j_e$ by using eqn.\eqref{coh_decomp}:
\be
\la \{z_e\}||\hat{j_f}||\{z_e\}\ra
=
\sum_{\{j_e\}}j_f\,
\prod_e\f{1}{(2j_e)!}
\,\la\{j_e,z_e\}||\{j_e,z_e\}\ra
\,.
\ee
Actually, we can generalize this formula to any observable $\cO(j_e)$ depending on the spin labels instead of the single observable $j_f$.
To study the behavior of the probability distribution in the $j_e$'s, we need the norm of the LS coherent intertwiners $\la\{j_e,z_e\}||\{j_e,z_e\}\ra$. These norms do not have a closed formula (up to now) despite detailed analysis \cite{more}, but we do have their asymptotic behavior obtained through a saddle point approximation \cite{LS}. Nevertheless, even without knowing the full exact behavior of this probability distribution, we can still extract some (minimal) information.

\begin{figure}[h]
\begin{center}
\includegraphics[height=40mm]{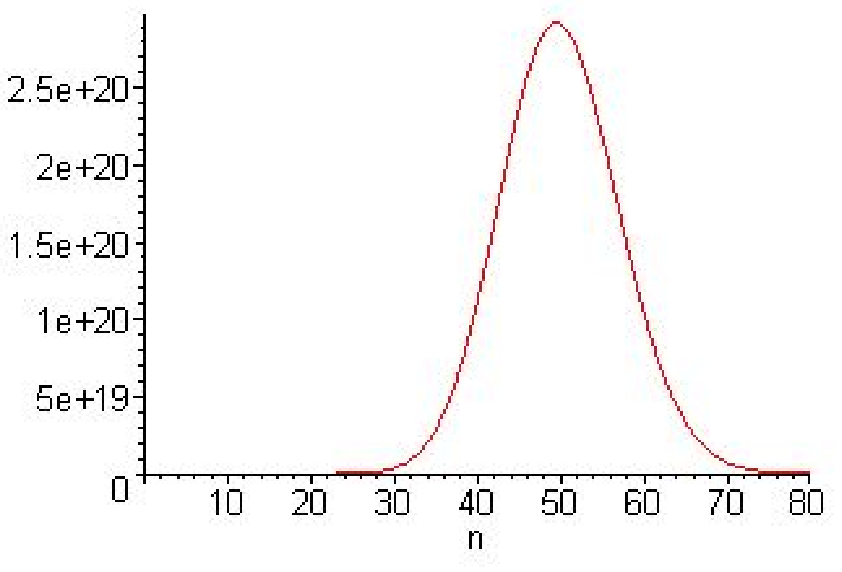}
\hspace*{5mm}
\includegraphics[height=40mm]{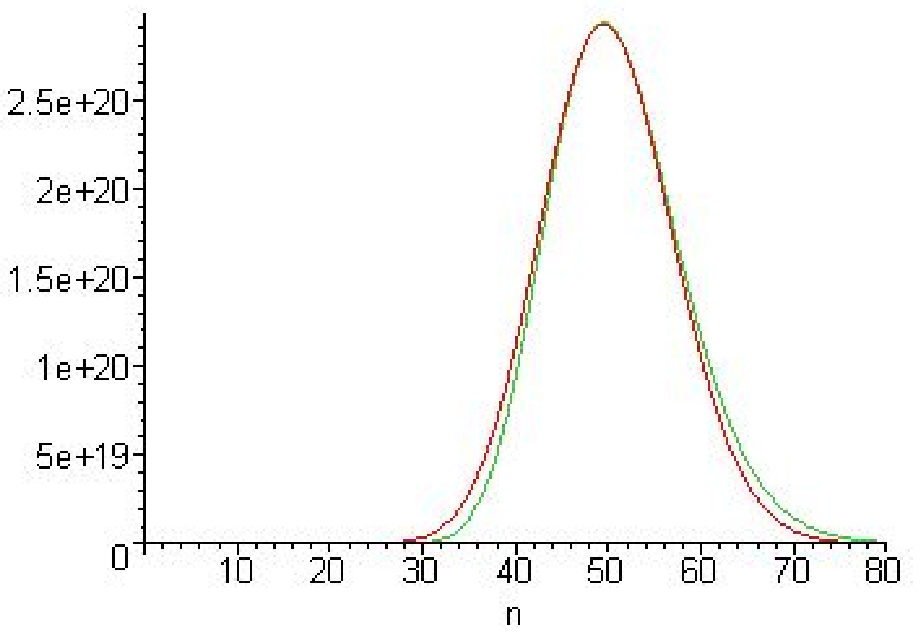}
\caption{On the left, plot of the Poisson distribution $P[j_f]$ describing the probability distribution of the spin label $j_f$ for a value $x=\la z_f|z_f\ra=50$. The x-axis gives the value of $2j_f$. On the right, its superposition with its Gaussian approximation around its maximum $2j_f\sim x=50$. \label{Poisson}}
\end{center}
\end{figure}

Indeed, the LS coherent intertwiner $||\{j_e,z_e\}\ra$ is the group averaged state of the tensor product of the $\SU(2)$ coherent states $|j_e,z_e\ra$ which are not normalized. The norm  of the $\SU(2)$ coherent states is simple $\la j_e,z_e|j_e,z_e\ra=\la z_e|z_e\ra^{2j_e}$. We can extract this norm for all the states. Then assuming that the LS coherent intertwiner defined as the group averaging of the tensor product of the normalized coherent states contains information about the coupling between the various legs of the intertwiner, the other factors can be considered as describing approximatively the probability distribution for each decoupled single spin label $j_f$:
\be
P[j_f]=\f{1}{(2j_f)!}\,\la z_f|z_f\ra^{2j_f}.
\ee
One directly recognizes a Poisson distribution. It is peaked about $2j_f=\la z_f|z_f\ra$ as expected. One can use the Stirling approximation for the factorial as before and check that (as well known):
\be
P[j_f]=\f{x^{2j_f}}{(2j_f)!}
\underset{j_f\gg1}{\sim} \f1{\sqrt{4\pi j_f}}e^{2j_f(1+\ln x-\ln 2j_f)}
\underset{2j_f\sim x}{\sim} \f{e^{2j_f}}{\sqrt{4\pi j_f}}e^{-\f1{4j_f}(2j_f-x)^2},
\qquad
\textrm{with}\quad
x=\la z_f|z_f\ra,
\ee
where the second approximation $\sim$ gives the Gaussian behavior of the Poisson distribution around its maximum, as shown on fig.\ref{Poisson}.

\subsection{Decomposing the Identity on the Intertwiner Space}

To conclude our summary of coherent intertwiner states, we would like to review the property that these coherent intertwiners provide us with an over-complete basis of the intertwiner space $\cH_N$. More exactly, the $\SU(2)$ coherent states $|j,z\ra$ form an over-complete basis of the space $\cV^j$ at fixed spin $j$, the LS coherent intertwiners $||\{j_e,z_e\}\ra$ form an over-complete basis of the space $\textrm{Inv}_{\SU(2)}\otimes_e \cV^{j_e}$ for fixed spins on all the legs, the $\U(N)$ coherent states $|J,\{z_e\}\ra$ form an over-complete basis of the space $\cR^J$ for fixed total area $J=\sum_e j_e$ (or equivalently fixed $\U(N)$ representation) and finally our coherent states $||\{z_e\}\ra$ span the whole space of $N$-valent intertwiners $\cH_N$.

Furthermore, we can write a decomposition of the identity on $\cH_N$ using our new coherent intertwiners. It is directly inherited from the decomposition of the identity on the Hilbert space $\cH^{HO}$ of a harmonic oscillator using the standard coherent states\footnotemark:
\be
\id_{\cH^{HO}}
\,=\,
\f1{\pi}\int d^2z\, e^{-|z|^2}\,|z\ra_{HO}{}_{HO}\la z|,
\qquad\textrm{with}\quad
|z\ra_{HO}\,\equiv\,
e^{z a^\dag}\,|0\ra
\,=\,
\sum_n \f{z^n}{\sqrt{n!}} |n\ra_{HO}
\ee
\footnotetext{
This identity is rather straightforward to show explicitly:
$$
\int d^2z\, e^{-|z|^2}\,\sum_{n,m} \f{z^n\bz^m}{\sqrt{n!m!}} |n\ra_{HO}{}_{HO}\la m|
\,=\,
\sum_n \int d^2z\, e^{-|z|^2}\, \f{(|z|^2)^n}{n!}|n\ra_{HO}{}_{HO}\la n|
\,=\,
\pi\,\sum_n |n\ra_{HO}{}_{HO}\la n|\,.
$$
}
Starting by applying this decomposition of the identity to the space $\cH^{HO}\otimes \cH^{HO}$ and projecting down to $\cV^j$ by fixing the total energy and thus the spin, we obtain\footnotemark:
\be
\label{decomp}
\id_{\cV^j}
\,=\,
\f1{\pi^2}\int d^4z\, e^{-\la z|z\ra}\,|j,z\ra\la j,z|\,.
\ee
\footnotetext{
It is possible to check directly this formula by decomposing the states $|j,z\ra$ on the basis $|j,m\ra$. We can also use the covariance property of the coherent states $|j,z\ra$ under the $\SU(2)$-action, which implies that the integral over $d^4z$ is proportional to the identity on $\cV^j$ by Sch\"ur lemma. Then we just need to check the trace of the operator:
$$
\int d^4z\, e^{-\la z|z\ra}\,\la j,z|j,z\ra
=
\int d^4z\, e^{-\la z|z\ra}\,\la z|z\ra^{2j}
=\pi^2\,(2j+1)!\,.
$$
}

Next, applying the decomposition of $\id_{\cH^{HO}}$ to the tensor product $\bigotimes_e^N (\cH^{HO}_e\otimes\cH^{HO}_e)$, we can decompose the identity using $2N$ copies of coherent states. Then we simply go down to the intertwiner space by group averaging $\cH_N\,=\, \textrm{Inv}_{\SU(2)}\bigotimes_e^N (\cH^{HO}_e\otimes\cH^{HO}_e)$ as explained in section\ref{irreps}-eqn.\eqref{int_space}. This provides an easy decomposition of the identity on $\cH_N$:
$$
\id_{\cH_N}
\,=\,\f1{\pi^{2N}}\int [d^4z_e]^N e^{-\sum_e \la z_e|z_e\ra}
\int_{\SU(2)} dg\,
g\vartriangleright\left(
e^{\sum_e z_e^0 a_e^\dag+ z_e^1 b_e^\dag}\,|0\ra
\la 0|\left(e^{\sum_e z_e^0 a_e^\dag+ z_e^1 b_e^\dag}\right)^\dag
\right),
$$
where $\SU(2)$ group elements act by conjugation. Taking into account the definition of the coherent intertwiners in \eqref{coh_decomp}, this simply means that our coherent intertwiners provide a decomposition of the identity:
\be
\id_{\cH_N}
\,=\,\f1{\pi^{2N}}\int [d^4z_e]^N e^{-\sum_e \la z_e|z_e\ra}
||\{z_e\}\ra\la\{z_e\}||\,.
\ee
From this decomposition of the identity, one can project it on spaces at fixed $J$ or further on spaces at fixed $\{j_e\}$ and derive the formulas in terms of the other coherent intertwiner states.

Here, we need to point out that we are integrating over all sets of spinors $\{z_e\}$ and not restricting ourselves to spinors satisfying the closure constraint $\sum_e |z_e\ra\la z_e|\propto \id$. Actually, all the  definitions of LS coherent intertwiners or $\U(N)$ coherent states or our new coherent intertwiners that we gave in the previous section do not depend on the closure constraints and work for generic spinors. The only things that change are the explicit expressions for the scalar products and norms. More precisely, for generic spinors, we have:
\beq
\la J,\{w_e\}|J,\{z_e\}\ra
&=&
\left(\f12\sum_{e,f} \la z_f|z_e][ w_e|w_f\ra\right)^J,\nn\\
\la J,\{z_e\}|J,\{z_e\}\ra
&=&
\left(\f12\sum_{e,f} |\la z_f|z_e]|^2\right)^{J}
=\left(\f14\left|\sum_e |\vV(z_e)|\right|^2-\f14\left|\sum_e \vV(z_e)\right|^2\right)^J\,.\nn
\eeq
When the closure constraint is satisfied, $\sum_e \vV(z_e)=0$, the norm of the state reduces to $A(z)^{2J}$ as before. We see that the more the closure constraint is violated, the more suppressed the norm of the coherent intertwiner is. This is similar to what happens with the LS coherent intertwiners \cite{LS}.

The important point about relaxing the closure constraint is that we can go in and out of it by a straightforward global $\SL(2,\C)$ action on the spinors \cite{un2,more}. $\SL(2,\C)$ transformations act as 2$\times$2 matrices on the spinors:
\beq
|z_e\ra & \rightarrow & \Lambda \vartriangleright |z_e\ra= |\Lambda z_e\ra,\nn\\
\sum_e |z_e\ra\la z_e |&\arr&
\Lambda\,\left(\sum_e |z_e\ra\la z_e|\right)\,\Lambda^\dag,\nn\\
{[}z_e|z_f\ra &\rightarrow&
[z_e|\Lambda^{-1}\Lambda |z_f\ra=[z_e|z_f\ra\,.\nn
\eeq
On the one hand, starting with an arbitrary set of spinors $z_e$, one can always choose a suitable $\Lambda\in\SL(2,\C)$ so that $\Lambda\,\left(\sum_e |z_e\ra\la z_e|\right)\,\Lambda^\dag\propto\id$.  On the other hand, the $\SU(2)$-invariant observables $[z_e|z_f\ra$ are furthermore invariant under the $\SL(2,\C)$-action, so that the coherent intertwiner states $|J,\{z_e\}\ra$ and $||\{z_e\}\ra$ are themselves invariant under the change $z_e\arr\Lambda z_e$ (for more details, the interested reader can refer to \cite{un2,un3}).

By gauging out this $\SL(2,\C)$-invariance, one can restrict the full integral over all spinors $z_e$ to an integral over sets of spinors satisfying the closure constraint. In order to do this consistently, one has to compute the corresponding Fadeev-Popov determinant. For the LS coherent intertwiners, this was investigated in \cite{more}. For the $\U(N)$ coherent states $|J,\{z_e\}\ra$ and our holomorphic coherent intertwiners $||\{z_e\}\ra$, this determinant is trivial due to the invariance of those states under $\SL(2,\C)$ \cite{un2,un3}.

\medskip

Now that we have consistently defined coherent intertwiner states and describe how they provide a decomposition of the identity on the intertwiner space, we will first explain how to glue them together to build coherent spin network states, then we will show how to use them in order to solve the holomorphic simplicity constraints at the quantum level.

\subsection{Coherent Spin Network States}


Let's generalize the previous construction of coherent intertwiners to full coherent spin network states on an arbitrary graph $\Gamma$.
We start with the classical spinor networks on the graph $\Gamma$ and consider the corresponding phase space parameterized by the spinors $z_e^v$ satisfying the closure constraints $\sum_{e\ni v} \vV(z_e^v)=0$ at every vertex and the gluing constraints $\vV(z_e^{s(e)})=\vV(z_e^{t(e)})$ on every edge, and invariant under the symmetries generated by those constraints, thus $\SU(2)$-transformations at every vertex and $\U(1)$-transformations on every edge. In short, our classical phase space on the graph $\Gamma$ is:
$$
(\C^4)^E//(\SU(2)^V\times\U(1)^E)\,,
$$
where $V$ and $E$ are respectively the number of vertices and edges of $\Gamma$. The symbol $//$ denotes the symplectic reduction. It amounts to both imposing the constraints and considering the orbits under the corresponding group action.

The standard spin network states provide an orthonormal basis of quantum state on this phase space. And we now would like to define coherent spin network states, which are peaked around each point of this phase space. Starting with a set of spinors on $\Gamma$, $\{z_e^v\}$ satisfying the closure and gluing constraints, we are going to use the previous definition of coherent intertwiners to define a coherent state on the whole $\Gamma$ graph labeled by this set of spinors.

To this purpose, we follow a straightforward logic and associate the coherent intertwiner $||\{z^v_e\}_{e\ni v}\ra$ to each vertex $v\in\Gamma$. Then we simply define our coherent quantum state on $\Gamma$ as the tensor product of these intertwiner states:
\be
\psi_{\{z_e^v\}}
\,=\,
\bigotimes_v ||\{z^v_e\}_{e\ni v}\ra\,.
\ee
These states are usually defined through their evaluation on group elements $\{g_e\}\in\SU(2)^E$:
\be
\psi_{\{z_e^v\}}(g_e)
\,=\,
\tr_e \bigotimes_e g_e \otimes \bigotimes_v ||\{z^v_e\}_{e\ni v}\ra\,.
\ee
To better understand the meaning of this evaluation on $\SU(2)$ group elements, we first expand our coherent intertwiners onto the LS coherent intertwiners. This re-introduces explicitly the $\SU(2)$ representation labels $j_e$. A subtlety is that this a priori leads to two spin labels $j_e^{s(e)}$ and $j_e^{t(e)}$ per edge for each of the two intertwiners living at the source and target vertices of the edge $e$. However, since we evaluate this expression on $\SU(2)$ group elements $g_e$, these two $\SU(2)$ representation labels must necessarily match $j_e^{s(e)}=j_e^{t(e)}$ and the evaluation is given in term of a single spin $j_e$ per edge of the graph \footnotemark.
\footnotetext{
A natural extension of the evaluation on $\SU(2)$ group elements is the evaluation on $\SL(2,\C)$ group elements. Indeed, as is well-known (e.g. see appendix of \cite{un0}), the Schwinger representation of $\SU(2)$ in terms of two harmonic oscillators also carries a unitary $\SL(2,\C)$ representation. Then we can evaluate our coherent spin network states on $\SL(2,\C)$ group elements, in which case the matching of the spins $j_e^{s(e)}$ and $j_e^{t(e)}$ one each  edge will be relaxed and we will have to describe the $\SL(2,\C)$-evaluation keeping these two spin labels per edge. \\
This natural extension to the evaluation on $\SL(2,\C)$ suggests a link between our coherent spin network states and the complexifier coherent states introduced earlier by Thiemann and al. \cite{thomas}. This should be investigated later.
}

Thus expanding the previous expression, we get:
\beq
\psi_{\{z_e^v\}}(g_e)
&=&
\sum_{\{j_e\}}\f{1}{\prod_e(2j_e)!}\, \tr_e \bigotimes_e D^{j_e}(g_e) \otimes \bigotimes_v ||\{j_e,z^v_e\}_{e\ni v}\ra \\
&=&
\int [dh_v]\,\sum_{\{j_e\}}
\prod_e \f{1}{(2j_e)!}\,
[j_e z_e^{s(e)}|h_{s(e)}^{-1}\,g_e\,h_{t(e)}|j_e z_e^{t(e)}\ra\,,
\eeq
where we have as before $|j,z]=|j,\varsigma z\ra=|z]^{\otimes 2j}$.
In this form, it is clear that this coherent state functional $\psi_{\{z_e^v\}}(g_e)$ is $\SU(2)$-invariant at every vertex $v$ and is fully holomorphic in the spinor labels $z_e^v$.

Moreover, this expression is clearly invariant under $\U(1)$-transformations on the spinors on each edge:
$$
z_e^{s(e)}\arr e^{i\theta_e}z_e^{s(e)},\qquad
z_e^{t(e)}\arr e^{-i\theta_e}z_e^{t(e)},
$$
and under $\SU(2)$-transformations on the spinors at each vertex since the coherent intertwiner states were themselves invariant. Thus our coherent spin network states $\psi_{\{z_e^v\}}$ are truly labeled by points (sets of spinors or spinor networks) in our constrained phase space $(\C^4)^E//(\SU(2)^V\times\U(1)^E)$.

Obviously, we have a sum over $\SU(2)$ spins $j_e$ and those are not fixed. We would like to point out that their distribution is not only fixed by the factor $1/(2j_e)!$ but also by the norm factors coming from the fact that the spinors are not normalized.

\medskip

Here, we have expanded our coherent spin network functional as a sum over spin labels $j_e$ because this is the usual way to discuss spin network states. However, in our case, the sum over the labels $j_e$ is straightforward due to the specific form of the coherent intertwiners and we have:
\beq
\psi_{\{z_e^v\}}(g_e)
\,=\,
\int [dh_v]\,
e^{\sum_e [ z_e^{s(e)}|h_{s(e)}^{-1}\,g_e\,h_{t(e)}|z_e^{t(e)}\ra}\,,
\eeq
where the matrix elements are all simply taken in the fundamental 2$\times$2 representation.

\subsection{Solving the Holomorphic Simplicity Constraints}

Now that we have reviewed the quantization of the classical spinor phase space for $\SU(2)$ intertwiners and spin networks, we can apply the construction of coherent intertwiners to the $\Spin(4)$ case and define simple coherent  intertwiners satisfying the holomorphic simplicity constraints.

As we have already seen earlier in section \ref{spin4_class}, since $\Spin(4)=\SU_L(2)\times\SU_R(2)$ factorizes exactly as the product of its two $\SU(2)$ subgroups, the classical phase space for $\Spin(4)$ intertwiners (and more generally spin networks) is just the tensor product of two copies of the classical phase space for $\SU(2)$ intertwiners (and spin networks). As a direct consequence, the quantization is straightforward: we define $\Spin(4)$-intertwiners as tensor products of two $\SU(2)$ intertwiners (one for the left $\SU(2)$ subgroup and the other for the right copy) and we obtain coherent $\Spin(4)$-intertwiners as tensor products of two coherent intertwiners:
$$
||\{j^L_e,z^L_e\}\ra\otimes||\{j^R_e,z^R_e\}\ra
\quad\textrm{and}\quad
|J^L,\{z^L_e\}\ra\otimes|J^R,\{z^R_e\}\ra
\quad\textrm{and}\quad
||\{z^L_e\}\ra\otimes||\{z^R_e\}\ra
\,,
$$
depending on the type of coherent intertwiners which we consider.

We are interested in writing the simplicity constraints at the quantum level and solving them using coherent intertwiner states. As we have seen in section \ref{simplicity_class}, the holomorphic simplicity constraints read at the classical level as:
$$
F^L_{ef}-\rho^2 F^R_{ef}=0\,,\quad \forall e\ne f\,.
$$
We define the quantum simplicity constraints as the direct quantization of these classical constraints:
\be
\hF^L_{ef}-\rho^2 \hF^R_{ef}=0\,,\quad \forall e\ne f\,.
\ee
It is easy to check that these constraints all commute with each other since the $\hF$-operators only involve annihilation operators (which comes from the fact that the $F$-observables are all holomorphic at the classical level). We call $\cHs$ the Hilbert space of intertwiners solving these constraints. Since our coherent intertwiners diagonalize the $\hF_{ef}$ operators, it is direct to give an overcomplete basis of simple coherent intertwiners for $\cHs$:
\be
|\{z_e\}\ra_\rho\,\equiv\,   ||\{\rho z_e\}\ra_L\otimes||\{z_e\}\ra_R,\qquad\qquad
(\hF^L_{ef}-\rho^2 \hF^R_{ef})\,|\{z_e\}\ra_\rho=0\,,
\ee
where the simplicity constraints are solved since $z^L_e=\rho z^R_e$ and thus $[z^L_e|z^L_f\ra=\rho^2\,[z^R_e|z^R_f\ra$. We have left the indices $L$ and $R$ on the quantum states to keep note of which intertwiner corresponds to $\SU_L(2)$ or to $\SU_R(2)$.

We can give the expectation values of the $\hE_{ef}$ operators, which measures the scalar product between the spinors on the legs $e$ and $f$ of the intertwiners, using the formula \eqref{Evalue} (for $\lambda=1$):
\be
\f{{}_\rho\la \{z_e\}|\hE^L_{ef}\,|\{z_e\}\ra_\rho}{{}_\rho\la \{z_e\}|\{z_e\}\ra_\rho}
\,=\,
\rho^2{\la z_e |z_f\ra}\,\f{\,I_2(2\rho^2A(z))}{I_1(2\rho^2A(z))}
\,=\,
\rho^2\,
\f{{}_\rho\la \{z_e\}|\hE^R_{ef}\,|\{z_e\}\ra_\rho}{{}_\rho\la \{z_e\}|\{z_e\}\ra_\rho}\,
\f{\,I_2(2\rho^2A(z))\,I_1(2A(z))}{I_1(2\rho^2A(z))\,I_2(2A(z))}\,,
\ee
where the last factor quickly converges to 1 when the area $A(z)$ grows large.
For the expectation values of further operators such as the scalar product operators, the interested reader can find more expressions in \cite{un3}.

In the following, we are going to use these new simple intertwiners, which solve the holomorphic simplicity constraints, to build simple coherent spin network states for the $\Spin(4)$ gauge group and to construct a new spinfoam model for (Riemannian) quantum gravity.

\subsection{$\U(N)$-Action on Simple Intertwiners}

A very interesting property of these simple coherent intertwiners is that they are covariant under a $\U(N)$ action, which can thus be used to deform them from one to another. This comes directly from the property of the $\SU(2)$ coherent intertwiners themselves \eqref{UNtransf1}:
\be
\label{UNtransf2}
\hat{U}\,|\{z_e\}\ra_\rho
\,=\,
|\{(Uz)_e\}\ra_\rho,\qquad
U=e^{i\alpha},
\quad
\hat{U}=e^{i\sum_{e,f}\alpha_{ef}(\hE^L_{ef}+\hE^R_{ef})}
=e^{i\sum_{e,f}\alpha_{ef}\hE^L_{ef}}e^{i\sum_{e,f}\alpha_{ef}\hE^R_{ef}}\,.
\ee
The generators of these $\U(N)$-transformations are the operators $\hE^L_{ef}+\hE^R_{ef}$.

It is actually interesting to check that the whole Hilbert space of simple intertwiners $\cHs$ is invariant under this $\U(N)$-action. Indeed, we compute the commutator of the $\u(N)$-generators with the simplicity constraints:
\be
[\hE^L_{ef}+\hE^R_{ef}\,,\,\hF^L_{gh}-\rho^2 \hF^R_{gh}]
\,=\,
\delta_{eh}(\hF^L_{fg}-\rho^2 \hF^R_{fg})-\delta_{eg}(\hF^L_{fh}-\rho^2 \hF^R_{fh})\,.
\ee
Thus if we start with a state $\psi$ satisfying the holomorphic simplicity constraints, $(\hF^L_{gh}-\rho^2 \hF^R_{gh})\,\psi=0$ for all pairs of legs $g,h$, and act on it with an operator $(\hE^L_{ef}+\hE^R_{ef})$ for an arbitrary pair of legs $e,f$, then the resulting state also satisfy the simplicity constraints for all pairs of legs. This shows that the space of simple intertwiners $\cHs$ carries a $\U(N)$-representation.

We hope that these $\U(N)$ transformations will later become useful to deform spin network states and study discrete diffeomorphisms of spinfoam amplitudes.

\subsection{Simple Spin Network States}

Similarly to how we build an overcomplete basis of coherent spin network states for $\SU(2)$ by assigning a coherent intertwiner state at each vertex of the graph, we can construct a basis of coherent spin network states for $\Spin(4)$. Here, we would like to focus on the construction of a basis of simple spin network states made of simple intertwiners which satisfy the holomorphic simplicity constraints at every vertex of the graph:
\be
\rpsi_{\{z_e^v\}}
\,\equiv\,
\bigotimes_v |\{z^v_e\}_{e\ni v}\ra_\rho\,.
\ee
These states are defined through their evaluation on group elements $\{G_e=(g^L_e,g^R_e)\}\in\Spin(4)^E$:
\beq
\rpsi_{\{z_e^v\}}(G_e)
&=&
\psi_{\{\rho z_e^v\}}(g^L_e)\,\psi_{\{z_e^v\}}(g^L_e)
\,=\,
\tr_e \bigotimes_e (g^L_e\otimes g^R_e) \otimes
\bigotimes_v ||\{\rho z^v_e\}_{e\ni v}\ra_L\otimes
||\{z^v_e\}_{e\ni v}\ra_R \\
&=&
\sum_{\{j^{L,R}_e\}}\f{1}{\prod_e(2j^L_e)!(2j^R_e)!}\,
\tr_e \bigotimes_e D^{j^L_e}(g^L_e)D^{j^R_e}(g^R_e)
\otimes \bigotimes_v ||\{j^L_e,\rho z^v_e\}_{e\ni v}\ra_L\otimes||\{j^R_e,z^v_e\}_{e\ni v}\ra_R \nn\\
&=&
\int [dh^{L,R}_v]\,\sum_{\{j^{L,R}_e\}}
\prod_e \f{1}{(2j^L_e)!(2j^R_e)!}\,
[j^L_e \rho z_e^{s(e)}|h^L_{s(e)}{}^{-1}\,g^L_e\,h^L_{t(e)}|j^L_e \rho z_e^{t(e)}\ra
[j^R_e z_e^{s(e)}|h^R_{s(e)}{}^{-1}\,g^R_e\,h^R_{t(e)}|j^R_e z_e^{t(e)}\ra \nn\\
&=&
\int [dh^{L,R}_v]\,\sum_{\{j^{L,R}_e\}}
\prod_e \f{\rho^{2j_L}}{(2j^L_e)!(2j^R_e)!}\,
 [j^L_e z_e^{s(e)}|h^L_{s(e)}{}^{-1}\,g^L_e\,h^L_{t(e)}|j^L_e z_e^{t(e)}\ra
[j^R_e z_e^{s(e)}|h^R_{s(e)}{}^{-1}\,g^R_e\,h^R_{t(e)}|j^R_e z_e^{t(e)}\ra\nn\\
&=&
\int [dh^{L,R}_v]\,
e^{\sum_e \rho^2[ z_e^{s(e)}|h^L_{s(e)}{}^{-1}\,g^L_e\,h^L_{t(e)}|z_e^{t(e)}\ra
+[ z_e^{s(e)}|h^R_{s(e)}{}^{-1}\,g^R_e\,h^R_{t(e)}|z_e^{t(e)}\ra}
\,.
\eeq

The big difference of the present proposal with the coherent intertwiner approach to the EPRL-FK spinfoam models \cite{FK,LS2,EPRL} based on the LS coherent intertwiner states \cite{LS} is that the EPRL-FK ansatz imposes strongly the diagonal simplicity constraints i.e. $j^L_e=\rho j^R_e$. This reduces our double sum over both $j^L_e$ and $j^R_e$ to a single sum over only the right (or left) spin labels. However, in our framework, we do not need to enforce the diagonal simplicity constraints by hand since, first, the holomorphic simplicity constraints implies the diagonal simplicity at the classical level, and second the spin labels are actually still peaked on the relation $j^L_e=\rho j^R_e$ but simply have a non-trivial spread around it.


\section{A New Spinfoam Model}

Since we have  introduced a new formulation for the simplicity constraints both at the classical and quantum levels and defined new coherent intertwiners and coherent spin networks that solved these holomorphic simplicity constraints, we would like to propose a new spinfoam model for 4d Riemannian quantum gravity constructed from these new coherent intertwiners. This can be considered as an improved version of the EPRL-FK spinfoam models \cite{EPR,FK,EPRL} based on a rigorous Gupta-Bleuer resolution of the simplicity constraints. As we will see, the definition of the model will be rather simple and we hope that the asymptotical analysis at large scale will similarly be simple.

As a first step, we will start by re-writing the spinfoam path integral for 4d BF theory with $\SU(2)$ as gauge group in terms of spinors and holomorphic coherent intertwiners. This is similar to the procedure which was started in \cite{LS} to write the discretized BF path integral in terms of the LS coherent intertwiners before imposing the simplicity constraints on the coherent intertwiners to derive the EPRL-FK spinfoam models \cite{FK,LS2}. Then our second step will be to impose our holomorphic simplicity constraints and define our new proposal for a spinfoam model.

\subsection{BF Theory in Terms of Spinors?}

Let's start by trying to define the topological spinfoam model for BF theory with gauge group $\SU(2)$. We will not review the spinfoam program here and we will assume that the reader is familiar with the spinfoam framework and structures. For reviews, the interested reader can refer to \cite{review}.

We will restrict ourselves for simplicity's sake to simplicial triangulations of the 4d space-time manifold with 4-simplices glued together along tetrahedra. Nevertheless, it is straightforward and obvious to generalize it to arbitrary cellular decomposition by considering the coherent intertwiners and the decomposition of the identity on the space of $N$-valent intertwiners for arbitrary $N$, and by defining the spinfoam amplitudes from the evaluation of the boundary spin networks for each 4-cell.

\medskip

\begin{figure}[h]
\begin{center}
\includegraphics[height=25mm]{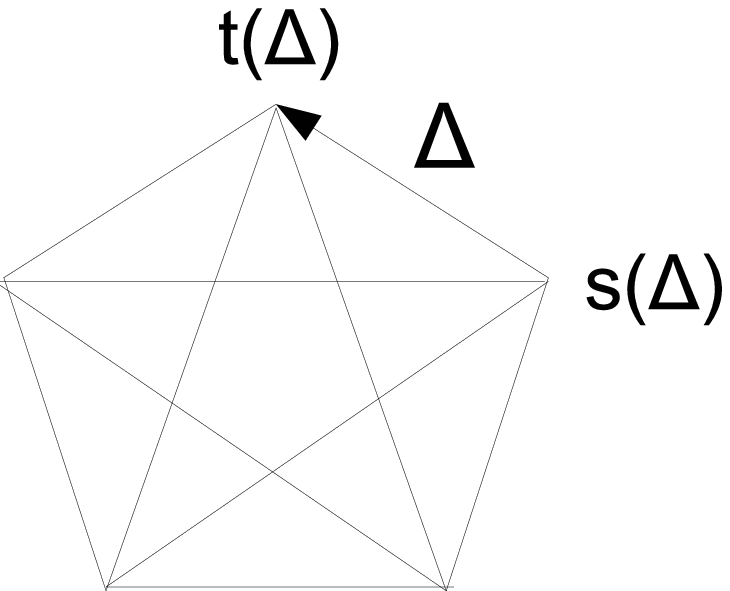}
\hspace*{10mm}
\includegraphics[height=25mm]{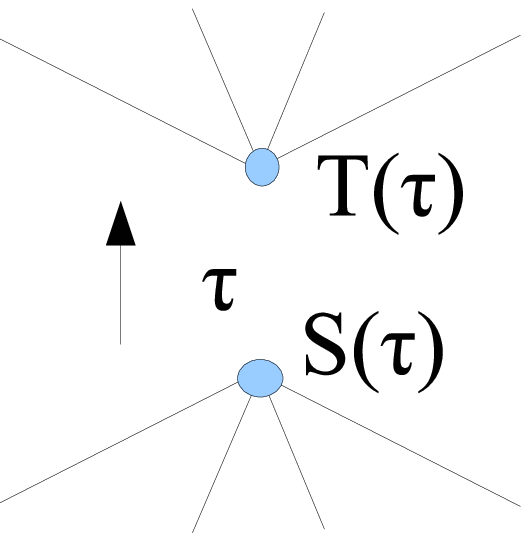}
\hspace*{10mm}
\includegraphics[height=25mm]{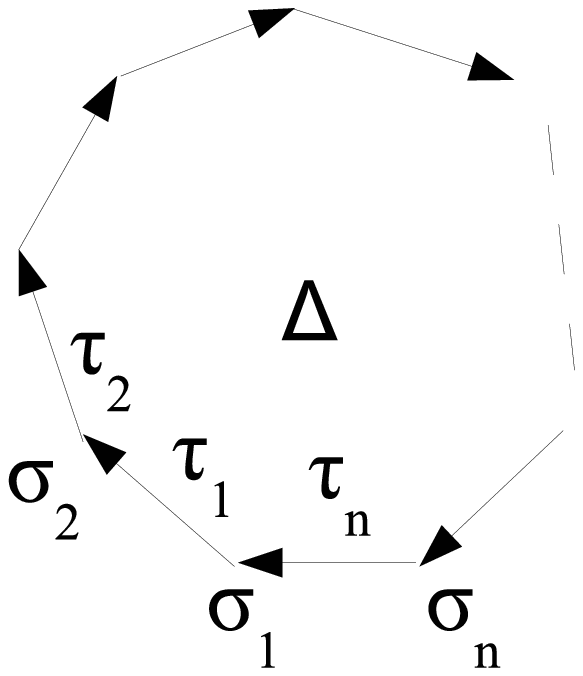}
\caption{From the left to the right: the boundary graph of a 4-simplex $\sigma$ (nodes are tetrahedra $\tau$ and links are triangles $\Delta$), a tetrahedron $\tau$ shared by the two 4-simplices $S(\tau)$ and $T(\tau)$, and a plaquette around a triangle $\Delta$ with all the 4-simplices and tetrahedra sharing the same triangle.\label{4simplex}}
\end{center}
\end{figure}

Thus let us consider a 4d triangulation made of 4-simplices. Spinfoams are made of two ingredients: a vertex amplitude which defines the local amplitude for the geometry of each 4-cell and an edge amplitude which defines how to glue the 4-cells together. The natural ansatz is:
\begin{itemize}

\item A {\it vertex amplitude} attached to each 4-simplex:

It defines the probability amplitude of the geometry of the 4-simplex. It is given by the evaluation of boundary spin network of the 4-simplex on the identity group element. The 4-simplex boundary graph is made of 5 nodes fully connected to each other, as shown on fig.\ref{4simplex}. Each of those nodes correspond to a tetrahedron of the 4-simplex. To each of the nodes or intertwiners, we attach a coherent intertwiner and build the resulting coherent spin network living on the 4-simplex boundary. Finally evaluating this spin network functional at the identity provides us with a amplitude depending on $2\times 10$ spinors:
\be
\cA_\sigma(z_\Delta^\tau)=
\psi_{\{z_\Delta^\tau\}}(\id)=
\tr\bigotimes_\tau||\{z_\Delta^\tau\}\ra=
\int [dh_\tau]^5\,e^{\sum_\Delta
[z_\Delta^{s(\Delta)}|h_{s(\Delta)}^{-1}h_{t(\Delta)}|z_\Delta^{t(\Delta)}\ra}\,,
\ee
where $\sigma$ denotes the 4-simplex, $\Delta$ and $\tau$ respectively label the ten triangles and five tetrahedra of $\sigma$. $s(\Delta)$ and $t(\Delta)$ respectively denote the source and target tetrahedra sharing the triangle $\Delta$ as depicted on fig.\ref{4simplex}. This vertex amplitude is completely holomorphic in the spinor variables $z_\Delta^\tau$.

\item An {\it edge amplitude} describing the gluing of the 4-simplices:

Each tetrahedron $\tau$ of the 4d triangulation glues two 4-simplices together. In each of these 4-simplices, an intertwiner is associated to this tetrahedron. For BF theory, the standard ansatz for the gluing is to insert the identity on the intertwiner space between the two 4-simplices. Using the decomposition of the identity in terms of coherent intertwiners, we have:
$$
\id_{\cH^\tau_{N=4}}
\,=\,
\f1{\pi^8}\int[d^4z_\Delta^\tau]^{\times 4}\,
e^{-\sum_{\Delta\in\tau}\la z^\tau_\Delta|z^\tau_\Delta\ra}\,
||\{z_\Delta^\tau\}\ra\la\{z_\Delta^\tau\}||\,.
$$
One thing to keep in mind is that each spinor variable $z_\Delta^\tau$ will enter the partition function twice, once holomorphic in $||\{z_\Delta^\tau\}\ra$ and once anti-holomorphic in $\la\{z_\Delta^\tau\}||$. To keep track of this, we have to introduce a new notation to decide for each tetrahedron $\tau$ which one of the two 4-simplices is the source $S(\tau)$ and which one is the target $T(\tau)$, as illustrated in fig.\ref{4simplex}. Then the vertex amplitude for the 4-simplex $\sigma=T(\tau)$ will be holomorphic in $z_\Delta^\tau$ while the vertex amplitude for $\sigma=S(\tau)$ will be anti-holomorphic in $z_\Delta^\tau$.

\end{itemize}

Putting these two ingredients together, we define a spinfoam partition for $\SU(2)$ BF theory expressed entirely in terms of coherent intertwiners and spinors. For a triangulated manifold $\cM$, we associate spinors $z_\Delta^\tau$ to each triangle in each tetrahedron and have auxiliary $\SU(2)$ group elements $h_\tau^\sigma$ associated to each tetrahedron in each 4-simplex:
\be
Z[\cM]
\,=\,
\int \prod_\tau\prod_{\Delta\in\tau}\f{d^4z_\Delta^\tau}{\pi^2}\,
e^{-\sum_{\Delta\in\tau}\la z^\tau_\Delta|z^\tau_\Delta\ra}\,
\prod_\sigma\int \prod_{\tau\in\sigma}dh_\tau^\sigma\,
e^{\sum_{\Delta\in\sigma}
[\varsigma^{\eps^{\sigma}_{s(\Delta)}}\, z_\Delta^{s(\Delta)}
|(h^\sigma_{s(\Delta)})^{-1}h^\sigma_{t(\Delta)}|
\varsigma^{\eps^{\sigma}_{t(\Delta)}}\,z_\Delta^{t(\Delta)}\ra}\,,
\ee
where the sign ${\eps^{\sigma}_{\tau}}$ is equal to 1 if the 4-simplex $\sigma$ is the source for the tetrahedron $\tau$ and is 0 if $\sigma=T(\tau)$.

This partition function is simply derived by associating to each tetrahedron a coherent intertwiner labeled by the appropriate spinors and gluing these intertwiners within each 4-simplex. A very interesting property of this spinfoam model is that it is directly defined by a discrete action principle. It could be interesting to compare it with other discrete action principle proposed for spinfoam models. Here, we will focus on showing that this partition function correctly defines the discrete path integral for topological BF theory.

We focus on a plaquette, dual to a given triangle $\Delta$. For simplicity's sake, we assume that it is consistently oriented all around the plaquette, as shown in fig.\ref{4simplex}:
$$
\left|
\begin{array}{l}
S(\tau_i)=\sigma_i, \\
T(\tau_i)=\sigma_{i+1},
\end{array}
\right.
\qquad\quad
\textrm{and in the 4-simplex }\sigma_i:\quad
\left|
\begin{array}{l}
s(\Delta)=\tau_{i-1}, \\
t(\Delta)=\tau_{i},
\end{array}
\right.\,,
$$
with the obvious identification $(n+1)\equiv 1$.
Dropping the subscript $\Delta$, the terms of the action involving the spinors all around the plaquettes are (taking care with the signs and relative orientations):
\be
\sum_{i=1}^n [z^{\tau_{i-1}}|(h^{\sigma_i}_{\tau_{i-1}}){}^{-1}h^{\sigma_i}_{\tau_i}|z^{\tau_i}]
\,=\,
\sum_{i=1}^n
\la z^{\tau_{i}}|(h^{\sigma_i}_{\tau_{i}}){}^{-1}h^{\sigma_i}_{\tau_{i-1}}|z^{\tau_{i-1}}\ra\,.
\ee
Then we can perform the integration over the spinor variables keeping in mind that each spinor $z$ enters the action twice, once as $|z\ra$ and once as $|z]$.
We only need the following Gaussian integral\footnotemark:
\be
\label{convo}
\f1{\pi^2}\int d^4z\, e^{-\la z|z\ra}e^{\la w| z\ra+\la z|\tw\ra}
\,=\,
e^{\la w|\tw\ra}\,.
\ee
%
%
\footnotetext{
We can either compute the Gaussian integral explicitly or use the decomposition \eqref{decomp} of the identity on $\cV^j$ in terms of coherent states:
\beq
\f1{\pi^2}\int d^4z\, e^{-\la z|z\ra}e^{\la w| z\ra+\la z|\tw\ra}
&=&
\f1{\pi^2}\int d^4z\, e^{-\la z|z\ra}
\sum_j\f{\la w| z\ra^{2j}}{(2j)!} \sum_k \f{\la z|\tw\ra^{2k}}{(2k)!}
\,=\,
\f1{\pi^2}\sum_j\f{1}{(2j)!^2}
\int d^4z\, e^{-\la z|z\ra}
\la j,w| j,z\ra\la j,z|j,\tw\ra\nn\\
&=&
\sum_j\f{1}{(2j)!}\la j,w| j,\tw\ra
\,=\,
e^{\la w|\tw\ra}.\nn
\eeq
}
Applying this to the integrals around the plaquette, we get:
$$
\f1{\pi^2}\int d^4z^{\tau_1}\,e^{-\la z^{\tau_1}|z^{\tau_1}\ra}
e^{\la z^{\tau_{2}}|h^{\sigma_2}_{\tau_{2}}{}^{-1}h^{\sigma_2}_{\tau_{1}}|z^{\tau_1}\ra+
\la z^{\tau_{1}}|h^{\sigma_1}_{\tau_{1}}{}^{-1}h^{\sigma_1}_{\tau_{n}}|z^{\tau_{n}}\ra}
\,=\,
e^{\la z^{\tau_{2}}|h^{\sigma_2}_{\tau_{2}}{}^{-1}h^{\sigma_2}_{\tau_{1}}
h^{\sigma_1}_{\tau_{1}}{}^{-1}h^{\sigma_1}_{\tau_{n}}|z^{\tau_{n}}\ra}.
$$
Calling $g_{\tau_i}\,\equiv\,h^{\sigma_{i+1}}_{\tau_{i}} (h^{\sigma_i}_{\tau_{i}}){}^{-1}$or more generally $g_\tau\equiv\,h_\tau^{T(\tau)}(h_\tau^{S(\tau)})^{-1}$, we can integrate this integration and finally obtain:
\be
\int \prod_i\f{e^{-\la z^{\tau_i}|z^{\tau_i}\ra}\,d^4z^{\tau_i}}{\pi^2}\,
e^{\sum_{i=1}^n
\la z^{\tau_{i}}|(h^{\sigma_i}_{\tau_{i}}){}^{-1}h^{\sigma_i}_{\tau_{i-1}}|z^{\tau_{i-1}}\ra}
\,=\,
\f1{\pi^2}\int d^4z\, e^{-\la z|z\ra}\,e^{\la z|G|z\ra}\,
\qquad\textrm{with}\quad
G=g_n..g_1
\ee
We can evaluate this last integral by expanding it into irreps of $\SU(2)$\footnotemark:
\be
\f1{\pi^2}\int d^4z\, e^{-\la z|z\ra}\,e^{\la z|G|z\ra}
=
\sum_j \f1{\pi^2\,(2j)!}\int d^4z\, e^{-\la z|z\ra}\,\la z|G|z\ra^{2j}
=
\sum_j \chi_j(G)
=\tdelta(G)\,,
\ee
\footnotetext{
We can change variables from the spinor $z$ to the 3-vector $\vV$, which would allow to see the relation between the modes $e^{\la z|G|z\ra}$ and the more usual functionals $e^{\tr (\vec{X}\cdot\vsigma) G}$ usually used in spinfoam constructions \cite{pr3,star}. Thus, using the change of integration measure derived in \eqref{measure}, we get:
$$
\f1{\pi^2}\int d^4z\, e^{-\la z|z\ra}\,e^{\la z|G|z\ra}
=
\f1{4\pi}\int \f{d^3\vV}{|V|}\,e^{-|V|}\,e^{\f12|V|\tr G}\,e^{\f12\vV\cdot\tr G\vsigma}.
$$
Introducing the parametrization of the group element $G$ as $G=\cos\theta\id+i\sin\theta\hat{u}\cdot\vsigma$ in terms of the class angle $\theta\in[0,2\pi]$ and the rotation axis $\hat{u}\in\cS^2$, we can separate the integration over the radius $|V|$ and the integration over the angular part $d^2\hat{V}$ and we get:
$$
\f1{\pi^2}\int d^4z\, e^{-\la z|z\ra}\,e^{\la z|G|z\ra}
=
\int_0^{+\infty} dV\,e^{-V(1-\cos\theta)}\,\f{\sin (V\sin\theta)}{V\sin\theta}
=\f{1}{2(1-\cos\theta)},
$$
where the last equality is a standard integral. This can be compared to the formula for the $\tdelta$-distribution as a sum over spin labels, if we compute the sum as a mere geometrical series:
$$
\sum_j \chi_j(g)=
\sum_{n\in\N^*} \f{\sin(n+1)\theta}{\sin\theta}
=\f1{2i\sin\theta}\left(\sum_{n\ge 1} e^{in\theta}-\sum_{n\ge 1} e^{-in\theta}\right)
=\f{\cos\f\theta2}{2\sin\theta\sin\f\theta2}
=\f1{2(1-\cos\theta)}.
$$
}
where we used the decomposition \eqref{decomp} of the identity on $\cV^j$ in terms of coherent states. The character $\chi_j(G)$ is by definition the trace of the matrix representing the group element $G$ in the representation of spin $j$. This distribution $\tdelta(G)$ should be compared to the $\delta$-distribution on $\SU(2)$:
\be
\delta(G)=
\sum_j (2j+1)\,\chi_j(G)
=
\f1{\pi^2}\int d^4z\, e^{-\la z|z\ra}\,\sum_j\f{(2j+1)}{(2j)!}\la z|G|z\ra^{2j}
=
\f1{\pi^2}\int d^4z\, e^{-\la z|z\ra}\,(1+\la z|G|z\ra)\,e^{\la z|G|z\ra}\,.
\ee
It is the factor $d_j=(2j+1)=\dim\cV^j$ that messes up the relation with the topological BF theory.

Indeed, performing all the integrations over the spinor variables and doing a change of variables from the $h_\tau^\sigma$ to the $g_\tau$, our spinfoam partition function in terms of coherent intertwiners is very similar to the standard discretized path integral for BF theory:
\be
Z[\cM]=\int \prod_\tau [dg_\tau]\,\prod_\Delta\tdelta\left(
\overrightarrow{\prod}_{\tau\ni\Delta}g_\tau
\right)\,,
\ee
but the difference resides in the fact that $\tdelta$ is not the $\delta$-distribution on $\SU(2)$. For instance, $\tdelta$ is not stable under convolution and, as a consequence, the partition function $Z[\cM]$ is not topological. Assuming that we haven't made any mistake in the normalization of the coherent states or in the measures of integration over the spinors or in the decomposition of the identity on the intertwiner space, we see a few possibilities to fix the issue of the $(2j+1)$-factor and recover BF theory:
\begin{itemize}

\item For each triangle, we can insert by hand the observable $(2j+1)$  on the link of a 4-simplex around the plaquette. This is done by inserting the operator $(E_{ee}+1)$ in the path integral where $e$ stands for the link corresponding to the triangle $\Delta$ within the chosen 4-simplex. The operator $E_{ee}$ is a differential operator in the relevant spinor $z$, which is simply $\la z|\pp_z\ra$. This method is straightforward to implement and gives the desired result. Nevertheless, it doesn't help us to understand where the $(2j+1)$-factor comes from.

\item It seems that the  $(2j+1)$-factor is the factor that enters the orthonormality of the matrix elements $D^j(g)$ of the Peter-Weyl theoreom for functions in $L^2(\SU(2))$. This would mean that we have to modify our edge amplitude and that we shouldn't insert directly the identity on the intertwiner space $\id_{\cH_N}$ but maybe insert a decomposition of the identity on $L^2(\SU(2)^{\times 4})$ instead. We haven't yet investigated how this can be implemented in terms of the spinor variables and there does not seem to be a natural alternative to the insertion of $\id_{\cH_N}$ between 4-cells.

\item Putting aside the coherent states and intertwiners and focusing on the discrete path integral defined in terms of spinors, another possibility is to modify the terms in the action $e^{\la \tz|g| z\ra}$ to $(\la \tz|g| z\ra+1)\,e^{\la \tz|g| z\ra}$ as suggested by the decomposition of $\delta(g)$ as an integral over spinors. This actually simply amounts to the insertion of a $(2j+1)$-factor or equivalently of the operator $(E_{ee}+1)$ on the corresponding wedge:
    $$
    e^{\la \tz|g| z\ra}=\sum_j \f{\la \tz|g| z\ra^{2j}}{(2j)!}
    \quad\longrightarrow\quad
    (\la \tz|g| z\ra+1)\,e^{\la \tz|g| z\ra}=\sum_j (2j+1)\,\f{\la \tz|g| z\ra^{2j}}{(2j)!}
    $$
    The problem is that the ``convolution" property of these modes is not nice. Indeed the equivalent of \eqref{convo} is now:
    $$
    \f1{\pi^2}\int d^4z\, e^{-\la z|z\ra}\,
    (\la w| z\ra+1)\,e^{\la w| z\ra}\,
    (\la z|\tw\ra+1)\,e^{\la z|\tw\ra}
    \,=\,
    (\la w|\tw\ra^2+3\la w|\tw\ra+1)\,e^{\la w|\tw\ra}\,.
    $$
    And the power of the factor in front of the exponential will increase as we integrate over the spinor variables around the plaquette. This simply means that we pick up extra $(2j+1)$-factors as we perform the integrations around the plaquette.

    A way out is to insert the factor $(\la w| z\ra+1)$ only once around the plaquette. Indeed we do have:
    $$
    \f1{\pi^2}\int d^4z\, e^{-\la z|z\ra}\,
    (\la w| z\ra+1)\,e^{\la w| z\ra}\,
    e^{\la z|\tw\ra}
    \,=\,
    (\la w|\tw\ra+1)\,e^{\la w|\tw\ra}\,.
    $$
    This means choosing one ``origin" 4-simplex for the plaquette and insert that factor there. This is exactly equivalent to the insertion of a $(2j+1)$-factor (or of the operator $(E_{ee}+1)$) on the corresponding wedge of the plaquette, which was our first proposed solution!

    Another way out would be to introduce a $\star$-product, which would deform the multiplication between modes so that they remain stable under convolution:
    $$
     \f1{\pi^2}\int d^4z\, e^{-\la z|z\ra}\,
    \left[(\la w| z\ra+1)\,e^{\la w| z\ra}\right]\,\star\,
    \left[(\la z|\tw\ra+1)\,e^{\la z|\tw\ra}\right]
    \,=\,
    (\la w|\tw\ra+1)\,e^{\la w|\tw\ra}\,.
    $$
    This is very similar to what happens when writing discrete action principle for BF theory in terms of local terms \cite{action_valentin,action_thomas,GFT3d}. Indeed, it turns out useful and more convenient to define the discretized path integral using the $\star$-product on $\R^3$ dual to the convolution product on $\SU(2)$ \cite{star}. It would actually be interesting to compare that $\star$-product previously introduced to the new $\star$-product between functions over spinors that we need here.

    Finally, instead of introducing a $\star$-product, maybe a suitable change of integration measure over the spinors could allow to realize the same procedure.

\end{itemize}

Finally, it seems that the most straightforward method to truly write the spinfoam amplitudes for BF theory is to insert by hand a factor $(\la \tz|g| z\ra+1)$ on one wedge (i.e. 4-simplex) of each plaquette. We can choose the ``origin" `for the plaquette for instance at $i=1$ . And this insertion simply amounts to the insertion of the operator $(E_{ee}+1)$, which produces the required factor $d_j=(2j+1)$ to turn the distribution $\tdelta(G)$ into $\delta(G)$. That way, we do recover an exact discretization of the  path integral for the topological BF theory.

We will investigate the other possibility of using a $\star$-product in the future and see if there is a way to write the exact discretized BF path integral in terms of the holomorpic coherent intertwiners without the insertions discussed above.

\medskip

We compare our new coherent state approach with the more standard method of expanding spinfoam amplitudes as sums over discrete spin labels. Besides the obvious disadvantage that the most natural spinfoam ansatz presented here doesn't exactly reproduce the topological partition function for $\SU(2)$ BF theory, it still has some promising aspects:

\begin{itemize}

\item Even if the most natural ansatz in our framework does not lead to the spinfoam amplitudes for BF theory, a slight modification with suitable (simple) observable insertions does allow us to recover the proper partition function.

\item The path integral is directly expressed in terms of coherent states and coherent intertwiners, which should simplify the study of the semi-classical limit.

\item We have exchanged the sum over spin labels, with integrals over complex variables. The path integral defined through integrals over coherent intertwiners can be directly written as a discretized action principle. This should simplify the study of the large scale asymptotics and the (semi-)classical regime of the amplitudes.

\item It is possible to expand explicitly the coherent intertwiners as sums over spin labels, through the exact formulas given in the earlier sections. These sums are more intricate than usual because spin labels are not a priori forced to be the same around a plaquette: given a triangle $\Delta$, we would have one spin $j_\Delta^\sigma$ for each 4-simplex $\sigma\ni\Delta$, i.e. for each wedge of the plaquette. It is the integrals over the spinors which allow to identify (or not) the wedge spins around the same plaquette.

\end{itemize}

\subsection{A New Holomorphic Vertex Amplitude}

Now that we have reformulated the spinfoam partition function for topological BF theory in terms of coherent intertwiners and spinors, we can propose our spinfoam model for 4d gravity with Riemannian signature.

To start with, we focus on the vertex amplitude associated to 4-cells of the triangulated manifold. In general, we will put a simple spin network on the boundary graph of the 4-cell, with coherent intertwiners solving the (holomorphic) simplicity constraints on the nodes and we will define the vertex amplitude as the the evaluation of the boundary spin network. For the sake of notational simplicity, we will focus on a simplicial triangulation made out of 4-simplices.

Then the boundary spin network of the 4-simplex is labeled by 2$\times$10 spinors living on each triangle in each tetrahedron, just as in the case of the pure $\SU(2)$ BF theory. The vertex amplitude for a 4-simplex $\sigma$ is then:
\beq
{}_\rho\cA_\sigma(z_\Delta^\tau)&=&
{}_\rho\psi_{\{z_\Delta^\tau\}}(\id)=
\psi_{\{\rho z_\Delta^\tau\}}(\id)\psi_{\{z_\Delta^\tau\}}(\id)=
\tr\bigotimes_\tau||\{z_\Delta^\tau\}\ra_\rho \nn\\
&=&
\int [dh_\tau]^5\,e^{\sum_{\Delta\in\sigma}
\rho^2[z_\Delta^{s(\Delta)}|h^L_{s(\Delta)}{}^{-1}h^L_{t(\Delta)}|z_\Delta^{t(\Delta)}\ra
[z_\Delta^{s(\Delta)}|h^R_{s(\Delta)}{}^{-1}h^R_{t(\Delta)}|z_\Delta^{t(\Delta)}\ra}
\,.
\eeq


Gluing these vertex amplitudes with the decomposition of the identity on the intertwiner space, we obtain the full spinfoam amplitude for a 4d triangulation, being careful of the relative orientations of tetrahedra and 4-simplices as in the previous section. This automatically provides us with a spinfoam amplitude given by the integrals over spinors $z_\Delta^\tau$ and auxiliary group elements $h^{\sigma\,L,R}_\tau$ of a discrete Lagragian. It will be very interesting in the future to compare this action principle with the other proposed discretized action for general relativity as a constrained BF theory \cite{action_valentin,action_thomas}.

The arbitrariness in our construction is the gluing of the vertex amplitude into a full spinfoam associated to the whole triangulation. Here, we chose the natural ansatz from the perspective of our spinorial construction, which is given by the insertion of the identity on the intertwiner space. However, as we have seen earlier, this is not the choice of edge amplitude that allows recovery of the  spinfoam amplitudes for topological BF theory. Nevertheless, for spinfoam models which are not topological invariant, the edge amplitude is not a priori not fixed and should be kept as an ambiguity in the definition of the model. It would be fixed a posteriori by the identification of a symmetry of the discrete partition function (such as discrete diffeomorphisms) and could change under coarse-graining (renormalization flow).


\section{Conclusion and Outlook}

Using the recently developed formulation of $\SU(2)$ group elements and functionals over $\SU(2)$ in terms of spinors, we have discussed the simplicity constraints (with Immirzi parameter) for discretized Riemannian 4d quantum gravity. Following the approach started in \cite{un3}, we have introduced a new set of holomorphic simplicity constraints. We have shown their equivalence at the classical level with the standard simplicity constraints. Then we have shown how to solve them using a new coherent intertwiners, which diagonalize the annihilation operators of the $\U(N)$ formalism for $\SU(2)$ intertwiners \cite{un1,un2,un3}. This truly realizes a quantization \`a la Gupta-Bleuer. Finally, we have explained how to glue these coherent intertwiners into coherent spin network states and defined a new spinfoam model for discretized Riemannian 4d quantum gravity whose boundary states solve the holomorphic simplicity constraints and whose amplitudes are given by the evaluation of the new coherent spin network states.

This new spinfoam model is formulated without reference to spin labels but directly through a discrete action principle and integrals over spinor variables. The diagonal simplicity constraints are not strongly enforced and we are no more restricted to simple irreducible representations of $\Spin(4)$. A possible side-effect is that this might allow a more detailed discussion of the possible renormalization and running of the Immirzi parameter in this spinfoam model.

This new model naturally opens the door to various questions:
\begin{itemize}
\item We should compare our new discrete Lagrangian to the other proposals for discretized Riemannian 4d quantum gravity.

\item We could study the asymptotics at large scale (large area) of the vertex amplitude of our new model and compare it to the asymptotics formula of the EPRL-FK models \cite{asymptEPRL}. It would provide us with a first check that the semi-classical behavior of our model is correct.

\item It would be interesting to see if the $\U(N)$ covariance of the coherent intertwiners can be turned into a $\U(N)$ symmetry for the spinfoam amplitude, in the hope of potentially understanding a action of discrete diffeomorphisms on our new spinfoam model.

\item It would also be interesting to investigate if our new vertex amplitude satisfies recursion relations, which would be written as differential equations in terms of the spinors. As it is understood that recursion relations are deeply linked to the topological/diffeomorphism invariance and dynamics of the spinfoam model  \cite{recursion1}, the hope is that such differential equations would reflect the dynamics and Hamiltonian constraints as was recently shown for BF theory \cite{recursion2}.

\item We should see if we can generate our new spinfoam amplitudes from a  suitable group field theory.

\item Since we are discussing the implementation of the simplicity constraints at the discrete level in spinfoams and we are relaxing them, it would be interesting to look at our new spinfoam models from the point of view of modified gravity theories defined from topological BF theory with relaxed simplicity constraints, such as bi-metric gravity theories as defined in \cite{bimetric}. Indeed, such modified gravity theories could arise at large scales from the renormalization of spinfoams.

\item It is necessary to generalize our approach to the Lorentzian case and build a spinfoam model for Lorentzian 4d quantum gravity. We need to investigate if we can have similar holomorphic simplicity constraints and coherent intertwiners. This is currently under investigation \cite{inprep_lorentz}.

\item Finally, we can also investigate the application of our spinorial framework and new spinfoam model to the recently introduced spinfoam cosmology \cite{sfcosmo}. It turns out that it simplifies both the formulation of the boundary data and the transition amplitudes, and allows us to see that the new spinfoam amplitudes satisfy a Hamiltonian constraint in this symmetry-reduced setting \cite{inprep-sfcosmo}.

\end{itemize}

\section*{Acknowledgments}

EL and MD were partially supported by the ANR ``Programme Blanc" grants LQG-09.
%

\appendix

\section{Spinors and Notations}

In this preliminary section, we introduce spinors and the related useful notations, following the previous works \cite{un3,un4,twistor}.

\subsection{Spinors}

Considering a spinor $z$,
$$
|z\ra=\mat{c}{z^0\\z^1}, \qquad
\la z|=\mat{cc}{\bar{z}^0 &\bar{z}^1},
$$
we define its dual spinor through  the duality map  $\varsigma$ acting:
\be
\varsigma\mat{c}{z^0\\ z^1}
\,=\,
\mat{c}{-\bar{z}^1\\\bar{z}^0},
\qquad \varsigma^{2}=-1.
\ee
This is an anti-unitary map, $\la \varsigma z| \varsigma w\ra= \la w| z\ra=\overline{\la z| w\ra}$, and we will write the dual spinor as
$$
|z]\equiv \varsigma  | z\ra,\qquad
[z| w]\,=\,\overline{\la z| w\ra}.
$$

We associate to the spinor $z\in\C^2$ a 3-vector $\vec{V}(z)\in\R^3$ defined from the projection of the $2\times 2$ matrix $|z\ra\la z|$ onto Pauli matrices $\sigma_a$ (taken Hermitian and normalized so that $(\sigma_a)^2=\id$):
\be \label{vecV}
|z\ra \la z| = \f12 \left( {\la z|z\ra}\id  + \vec{V}(z)\cdot\vec{\sigma}\right).
\ee
The norm of this vector is $|\vec{V}(z)| = \la z|z\ra= |z^0|^2+|z^1|^2$  and its components are given explicitly as:
\be
V^z=|z^0|^2-|z^1|^2,\qquad
V^x=2\,\Re\,(\bar{z}^0z^1),\qquad
V^y=2\,\Im\,(\bar{z}^0z^1).
\ee
The spinor $z$ is entirely determined by the corresponding 3-vector $\vec{V}(z)$ up to a global phase. We can give the inverse map:
\be
z^0=e^{i\phi}\,\sqrt{\f{|\vec{V}|+V^z}{2}},\quad
z^1=e^{i(\phi-\theta)}\,\sqrt{\f{|\vec{V}|-V^z}{2}},\quad
\tan\theta=\f{V^y}{V^x},
\ee
where $e^{i\phi}$ is an arbitrary phase.

Then the map $\varsigma$ sends the 3-vector $\vec{V}(z)$ onto its opposite:
\be
|z][ z| = \f12 \left({\la z|z\ra}\id - \vec{V}(z)\cdot\vec{\sigma}\right).
\ee

\subsection{Change of Integration Variables}

Since the spinor $z$ is entirely determined by the 3-vector $\vV$ and a phase $\phi$, we can compute the change of integration variable from $d^4z$ to a measure $d^4\mu(\vV,\phi)$. Actually it is more interesting to consider functions of the spinor $z$ which do not depend on the phase $\phi$, for instance functions of the matrix $|z\ra\la z|$. In this case, we can show that:
\be
\label{measure}
\f1{\pi^2}\int d^4z\,e^{-\la z|z\ra}\,f(\vV(z))
\,=\,
\f1{4\pi}\int \f{d^3\vV}{|\vV|}\,e^{-|\vV|}\,f(\vV)\,.
\ee
It is straightforward to prove by assuming that the measure on $\vV$ should be invariant under 3d rotations and then evaluating it over the basis of functions $|\vV|^n=\la z|z\ra^{n}$ of functions invariant  3d rotations.

\subsection{Closure of $N$ Spinors}

Considering the setting necessary to describe intertwiners with $N$ legs, we consider $N$ spinors $z_e$ and their corresponding 3-vectors $\vV(z_e)$.

We require that the $N$ spinors satisfy a closure
condition, i.e  that the sum of the corresponding 3-vectors
vanishes, $\sum_e \vec{V}(z_e)=0$. Coming back to the definition of
the 3-vectors $\vV(z_e)$, the closure condition is easily translated
in terms of $2\times 2$ matrices as the condition $\sum_e |z_e\ra \la z_e| \propto \id$:
\be
\sum_e |z_e\ra \la z_e|=A(z)\id,
\qquad\textrm{with}\quad
A(z)\equiv\f12\sum_e \la z_e|z_e\ra=\f12\sum_e|\vec{V}(z_e)|.
\ee
This further translates into quadratic constraints on the spinors:
\be
\sum_e z^0_e\,\bar{z}^1_e=0,\quad
\sum_e \left|z^0_e\right|^2=\sum_e \left|z^1_e\right|^2=A(z).
\label{closure}
\ee
In simple terms, it means that the two components of the spinors, $z^0_e$ and $z^1_e$, are orthogonal $N$-vectors of equal norm.

\subsection{Spinors and $\SU(2)$ Group Elements}

Given two spinors, $|z\ra$ and  $|w\ra$, there exists a {\it unique} group element $g\in\U(2)$ which maps one onto the other, i.e such that:
\be
g\f{|z\ra}{\sqrt{\la z|z\ra}}
\,=\,
\f{|w\ra}{\sqrt{\la w|w\ra}},
\qquad
g^\dag g=\id\,.
\ee
Its explicit expression in terms of the spinors is:
\be
g\,=\,
\f{|w\ra\la z|+|w][ z|}{\sqrt{\la z|z\ra\,\la w|w\ra}}.
\ee
It is direct to realize that this defines an $\SU(2)$ group element by checking that $g^\dag g=\id$ and $\tr g\in\R$.

\section{Observables $E$ and $F$: Poisson brackets and Commutation relations}
\label{app_commEF}

The Poisson brackets of the $\SU(2)$-observables are:
\bes
{\{}E_{ef},E_{gh}\}&=&
-i\left(\delta_{fg}E_{eh}-\delta_{eh}E_{gf} \right)\nn\\
{\{}E_{ef},F_{gh}\} &=& -i\left(\delta_{eh}F_{fg}-\delta_{eg}F_{fh}\right),\qquad
{\{}E_{ef},\bF_{gh}\} = -i\left(\delta_{fg}\bF_{eh}-\delta_{fh}\bF_{eg}\right), \\
{\{} F_{ef},\bF_{gh}\}&=& -i\left(\delta_{eg}E_{hf}-\delta_{eh}E_{gf} -\delta_{fg}E_{he}+\delta_{fh}E_{ge}\right), \nn\\
{\{} F_{ef},F_{gh}\} &=& 0,\qquad {\{} \bF_{ef},\bF_{gh}\} =0.\nn
\ees
At the quantum level, these observables become $\SU(2)$-invariant operators:
\bes
\hE_{ef}&=&a^\dag_e a_f+ b^\dag_e b_f, \nn\\
\hF_{ef}&=&a_e b_f- b_e a_f, \nn \\
\hFd_{ef}&= &a^\dag_e b^\dag_f- b^\dag_e a^\dag_f \nn.
\ees
They form a closed algebra, which mirrors the Poisson algebra given above:
\bes
{[}\hE_{ef},\hE_{gh}]&=&
\delta_{fg}\hE_{eh}-\delta_{eh}\hE_{gf} \nn\\
{[}\hE_{ef},\hF_{gh}] &=& \delta_{eh}\hF_{fg}-\delta_{eg}\hF_{fh},\qquad
{[}\hE_{ef},\hFd_{gh}] = \delta_{fg}\hFd_{eh}-\delta_{fh}\hFd_{eg}, \\
{[} \hF_{ef},\hFd_{gh}] &=& \delta_{eg}\hE_{hf}-\delta_{eh}\hE_{gf} -\delta_{fg}\hE_{he}+\delta_{fh}\hE_{ge}
+2(\delta_{eg}\delta_{fh}-\delta_{eh}\delta_{fg}), \nn\\
{[} \hF_{ef},\hF_{gh}] &=& 0,\qquad {[} \hFd_{ef},\hFd_{gh}] =0.\nn
\ees


\end{document}